\documentclass[twocolumn]{aastex62}
\usepackage{natbib}
\usepackage{graphicx}
\usepackage{float}

\usepackage[caption=false]{subfig}

\usepackage{xcolor}
\usepackage{xspace}
\usepackage{booktabs}
\usepackage{verbatim}
\graphicspath{{./}{figs/}}

\received{2019 March 27}
\revised{2019 July 24}
\accepted{2019 July 30}

\shorttitle{Dynamical Rings in Protoplanetary Disk Simulations}
\shortauthors{Kadam et al.}

\newcommand*{\eg}{e.g.\@\xspace}
\newcommand*{\Eg}{E.g.\@\xspace}
\newcommand*{\ie}{i.e.\@\xspace}
\newcommand{\simname}[1]{\texttt{#1}}

\newcommand{\bl}[1]{\mbox{\boldmath$ #1 $}}

\makeatletter
\newcommand*{\etc}{%
    \@ifnextchar{.}%
        {etc}%
        {etc.\@\xspace}%
}

\begin{document}

\title{Dynamical Gaseous Rings in Global Simulations of Protoplanetary Disk Formation}

\correspondingauthor{Kundan Kadam}
\email{kundan.kadam@csfk.mta.hu}

\author[0000-0002-0786-7307]{Kundan Kadam}
\affiliation{Konkoly Observatory, Research Center for Astronomy and Earth Sciences, Hungarian Academy of Sciences, Konkoly-Thege Mikl\'{o}s \'{u}t 15-17, 1121 Budapest, Hungary}

\author{Eduard Vorobyov}
\affiliation{Research Institute of Physics, Southern Federal University, Stachki Ave. 194, 344090 Rostov-on-Don, Russia}
\affiliation{Department of Astrophysics, The University of Vienna, A-1180 Vienna, Austria}

\author{Zsolt Reg\'{a}ly}
\affiliation{Konkoly Observatory, Research Center for Astronomy and Earth Sciences, Hungarian Academy of Sciences, Konkoly-Thege Mikl\'{o}s \'{u}t 15-17, 1121 Budapest, Hungary}

\author{\'{A}gnes K\'{o}sp\'{a}l}
\affiliation{Konkoly Observatory, Research Center for Astronomy and Earth Sciences, Hungarian Academy of Sciences, Konkoly-Thege Mikl\'{o}s \'{u}t 15-17, 1121 Budapest, Hungary}
\affiliation{Max Planck Institute for Astronomy, K\"onigstuhl 17, D-69117 Heidelberg, Germany}

\author{{P}\'{e}ter \'{A}brah\'{a}m}
\affiliation{Konkoly Observatory, Research Center for Astronomy and Earth Sciences, Hungarian Academy of Sciences, Konkoly-Thege Mikl\'{o}s \'{u}t 15-17, 1121 Budapest, Hungary}



\begin{abstract}
Global numerical simulations of protoplanetary disk formation and evolution were conducted in thin-disk limit, where the model included magnetically layered disk structure, a self--consistent treatment for the infall from cloud core as well as the smallest possible inner computational boundary. We compared the evolution of a layered disk with a fully magnetically active disk. We also studied how the evolution depends on the parameters of the layered disk model --- the MRI triggering temperature and active layer thickness --- as well as the mass of the prestellar cloud core. 

With the canonical values of parameters a dead zone formed within the inner $\approx${15} au region of the magnetically layered disk. The dead zone was not a uniform structure and long--lived, axisymmetric, gaseous rings ubiquitously formed within this region {due to the action of viscous torques.} The rings showed a remarkable contrast in the disk environment as compared to a fully magnetically active disk and were characterized by high surface density and low effective viscosity.  Multiple gaseous rings could form simultaneously in the dead zone region which were highly dynamical and showed complex, {time--dependent} behavior such as inward migration, {vortices, gravitational instability and large--scale spiral waves}. An increase in MRI triggering temperature had only marginal effects, while changes in active layer thickness as well as the initial cloud core mass had significant effects on the structure and evolution of the inner disk. Dust with large fragmentation barrier could be trapped in the rings, which may play a key role in planet formation.

\end{abstract}



\section{Introduction} 
\label{sec:intro}
As an inevitable consequence of angular momentum conservation, a low-mass star is formed within a collapsing cloud core via accretion through a protoplanetary disk.
The inner regions of such protoplanetary disks are of great interest to astronomers, since their evolution is intimately related to the formation of planetary systems.
In order to be able to accrete mass onto the central star, the protoplanetary disk needs get rid of the angular momentum of the accreting material.
The accretion process is believed to be inherently time-dependent as conjectured from the ``luminosity problem" in young stellar objects, as the observations of young clusters consistently show about an order of magnitude less luminosity as compared to the theoretical models \citep[\eg][]{Eisner2005,Dunham2010}. 
The existence of young eruptive objects, i.e. FUor and EXor variables which undergo sudden accretion events, also suggests that a significant amount of stellar mass is accreted during episodic accretion outbursts \citep[\eg][]{HK1996,DV2012}. 
Although substantial progress has been made over the last few decades, the details of the processes involved in the angular momentum transport in protoplanetary disks remain elusive.

One of the major insights into the accretion in a protoplanetary disk was the magnetically layered disk structure proposed by \cite{Gammie1996}.
{A disk needs to have sufficient levels of ionization to maintain angular momentum and mass transport through magnetorotational instability (MRI) \citep{HGB1995}. 
 The Galactic cosmic rays are thought to be the predominant source of ionization which penetrate an approximately constant column density, ${\rm \Sigma_{a}}$, at the disk surfaces \citep{UN1981}.
The stellar far ultraviolet and X-ray radiation combined have about an order of magnitude smaller effects at a distance of 1 au from the star \citep{Bergin2007, PBC2011,IG1999}. 
The radioactive decay of short--lived isotopes can be an additional source of ionization throughout the disk, however, the ionization rate they can provide is very small \citep{UN2009, Cleeves2013}.
Thus a typical protoplanetary disk forms a sandwich structure with a magnetically inactive ``dead zone" near the midplane at a few au, and a partially-ionized, magnetically ``active layer" at its surface.} 
As a natural consequence of this bottleneck, the material coming from the outer disk accumulates in the vicinity of the dead zone, and this can cause the inner disk to be susceptible to instabilities. 
In the presence of heating sources, the midplane temperature in the dead zone can rise and trigger ionization at a certain temperature (${\rm T_{crit}}$), \eg, because of ionization of alkali metals, or evaporation of small dust grains.
Thus MRI turbulence can be triggered in the inner disk regions and the sudden increase in viscosity would result in the accretion of the accumulated material onto the star. 
This is one of the mechanisms through which the protoplanetary disk accretion can be inherently episodic \citep[][]{Armitage2001, Zhu2010a, Martin2012, Bae2014}.

A number of numerical studies have used the magnetically layered disk model as a basis to demonstrate time-dependent accretion in protoplanetary disks.
Majority of these investigations were conducted with one dimensional disk models, which use local approximation of the angular momentum transport through self--gravity \citep[\eg][]{Armitage2001, Zhu2010a, Martin2012}. 
In reality a self--gravitating disk will exhibit intrinsically asymmetrical and non-local processes such as fragmentation, spirals and clumps.
Thus only a general insight into the protoplanetary disk evolution could be gained.
{An insight into the physics of the dead zone can also be gained by considering detailed microphysics and non-ideal magnetohydrodynamic effects, typically calculated within a local shearing-box approximation 
\citep[][]{BS2013, DT2015, Lesur2014}.
Multi--dimensional, global numerical investigations have their own limitations. 
Only short term evolution is feasible with three dimensional (or two dimensional, axisymmetric) disk simulations \citep[\eg][]{Zhu2009, Dzyurkevich2010, Ruge2016, Bai2017}, while other studies such as \cite{Lyra2009, Bae2014} use ad hoc prescription of the initial state of the disk and the mass infall from the parent cloud core.
The precise, long-term evolution of the protoplanetary disk and its dead zone would inevitably depend on the initial conditions in the prestellar core, which need to be considered self--consistently.
}       

In this paper we investigate the evolution of the inner regions of protoplanetary disks with respect to the layered accretion theory. 
We performed global hydrodynamic simulations of protoplanetary disks in the thin-disk limit, starting with the collapse phase of the cloud core, so that the initial conditions are carefully taken into account. 
The layered disk structure was emulated by an adaptive prescription of effective $\alpha$-parameter, which was calculated as a linear combination of $\alpha$ in the active and the dead region, weighted with the respective gas surface densities. 
In accordance with recent studies \citep[\eg][]{Turner2007,Okuzumi2011}, we have also considered a small but non-negligible dead zone viscosity which is driven by the MRI turbulence in the active layers.
The smallest possible inner computational boundary was considered with an implementation of inflow-outflow boundary condition, which allowed us to study the evolution of the inner disk at a sub-au distance scale.

As expected, we found that the dead zone developed in the layered disk simulations, however within this region high surface density gaseous rings formed.
{Maxima in surface density and pressure are shown to form near steep transition in the viscosity profile at the boundaries of the dead zone due to the mismatch in mass transfer rate, while 
disk substructure, gaps and rings have been the subject of several theoretical studies \citep[\eg][]{Chatterjee2014, Dzyurkevich2010, Flock2015, Guilera2017,Lyra2009, Ruge2016}.}
Here we investigate the formation and evolution of the {gaseous} rings as well as the dead zone in a self--consistent manner, and show that the structure and behavior of the inner disk is much more complex.
The detailed study of the magnetically layered disk, especially with respect to the local conditions in the rings, could be crucial for understanding the processes involved in planet formation.
A fully MRI active disk with a constant $\alpha$-parameter is widely used for modeling angular momentum transport observed in protoplanetary disks \citep[\eg][]{Hartmann1998,McKee2007}. 
In order to demonstrate the contrast, especially in the inner disk region, we made detailed comparison of the evolution of a fiducial magnetically layered disk with that of a fully MRI active disk.
We also studied the effects of the variation of the layered disk parameters --- ${\rm T_{crit}}$ and ${\rm \Sigma_{a}}$ --- as well as the mass of the gas in the initial cloud core.

The structure of this paper is as follows. In Section \ref{sec:methods} we describe our methods --- the numerical model, computational boundary and the initial conditions --- and in Section \ref{sec:results} we present our results. We elaborate on the defining characteristics of the gaseous rings in the fiducial layered disk model in contrast with the analogous fully MRI active disk in Section \ref{subsec:properties}. In Section \ref{subsec:dynamical} we concentrate on the evolution of the layered disk and the dynamical aspects of the rings. The implications of these rings on planet formation are considered in Section \ref{subsec:planets}. 
The effects of the model parameters on the disk structure are described in Section \ref{subsec:parameters}, and we summarize our conclusions in Section \ref{sec:summary}.

\section{Methods}
\label{sec:methods}

Numerical hydrodynamics simulations of disk formation and long-term evolution present a computational challenge. They have to capture spatial scales from sub-au to thousands of au with a numerical resolution that is sufficient to resolve disk substructures on au-scales (e.g., gaseous clumps formed via gravitational fragmentation). Since most of our knowledge about disk characteristics is derived from observations of T Tauri disks, numerical codes need to follow disk evolution from the embedded, optically obscured phase to the optically revealed T~Tauri phase. This implies evolution times on the other of a million years.
These requirements make fully three-dimensional simulations prohibitively expensive. Even simpler 3D models with a larger sink cell ($\ge$~several~au) and shorter integration times ($\le 0.1$~Myr) require enormous computational resources --- thousands of CPU cores and millions of CPU hours --- not easily available for the typical institutional infrastructure.

This is the reason why simpler numerical models of disk formation and evolution are still widely used to derive disk characteristics for a wide parameter space. The most often used approach is to employ various modifications of the one-dimensional viscous evolution equation for the surface density of gas as introduced in \citet{Pringle1981}. 
The general drawback of this approach is the inherent lack of azimuthal disk sub-structures (such as spiral arms, gaseous clumps, vortices), which can play a substantial role in the global structuring of protostellar/protoplanetary disks, as well as in their mass and angular momentum transport.
Moreover, the underlying assumption of negligible gas pressure makes dust evolution simulations within this model inconsistent. 

These limitations are relaxed in the disk models employing numerical hydrodynamics equations in the thin--disk limit. Much in the same way as the one-dimensional viscous evolution equation of \citet{Pringle1981}, these models assume a negligible impact of vertical motions on the global evolution of a circumstellar disk, expressed in the adoption of the local hydrostatic equilibrium. These models also assume a small vertical disk scale height with respect to the radial position in the disk, allowing for the integration of main hydrodynamic quantities in the vertical direction and the use of these integrated quantities in the hydrodynamics equations.

The obvious advantage of the thin-disk models is that they, unlike the viscous evolution equation, employ the full set of hydrodynamic equations, being at the same time computationally inexpensive in comparison to the full three-dimensional approach. The non--axisymmetric disk sub--structures (such as spiral arms, clumps, vortices) can be realistically modeled at least within the simplifying assumptions of the thin-disk model, which is impossible in the approach of \citet{Pringle1981}. The disadvantage of the thin-disk models is the lack of the disk vertical structure, although models now start to emerge that relax this limitation  and allow for the on-the-fly reconstruction of the disk vertical structure \citep{VP2017}.  Although intrinsically limited in its ability to model the full set of physical phenomena that may occur in star and planet formation, the thin-disk models nevertheless present an indispensable tool for studying the long-term evolution of protoplanetary disks for a large number of model realisations with a much higher realism than is offered by the simple one-dimensional viscous disk models.
\subsection{Numerical model}
\label{subsec:model}

The equations of mass, momentum, and energy transport in the thin-disk limit are:
\begin{equation}
\frac{\partial\Sigma}{\partial t}=-\bl{\nabla}_{p}\cdot\left(\Sigma \bl{v}_{p}\right),\label{eq:mass}
\end{equation}

\begin{equation}
\frac{\partial}{\partial t}\left(\Sigma \bl{v}_{p}\right)+\left[\bl{\nabla} \cdot \left(\Sigma 
\bl{v}_{p}\otimes \bl{v}_{p}\right)\right]_{p}=-\bl{\nabla}_{p} P+\Sigma \bl{g}_{p}+\left(\bl{\nabla}\cdot
\bl{\Pi}\right)_{p},\label{eq:momentum}
\end{equation}

\begin{equation}
\frac{\partial e}{\partial t}+\bl{\nabla}_{\mathrm{p}}\cdot\left(e v_{p}\right)=-P\left(\bl{\nabla}_{p} \cdot v_{p}\right)-\Lambda+\Gamma+\left(\bl{\nabla} \cdot v\right)_{pp'}: \bl{\Pi}_{pp'},
\label{eq:energy}
\end{equation}
where subscripts $p$ and $p'$ refer to the planar components $(r,\phi)$
in polar coordinates, $\Sigma$ is the mass surface density, $e$
is the internal energy per surface area, $P$ is the vertically integrated
gas pressure calculated via the ideal equation of state as $P=(\gamma-1)e$
with $\gamma=7/5$, $\bl{v}_{p}=v_{\mathrm{r}}\hat{\bl{r}}+v_{\mathrm{\phi}}\hat{\bl{\phi}}$
is the velocity in the disk plane, $\bl{g}_{p}=g_{\mathrm{r}} \hat{\bl{r}} 
+ g_{\mathrm{\phi}} \hat{\bl{\phi}}$
is the gravitational acceleration in the disk plane and 
$\bl{\nabla}_{\mathrm{p}}=\hat{\bl{r}}\partial/\partial r+\hat{\bl{\phi}}r^{-1}\partial/\partial\phi$
is the gradient along the planar coordinates of the disk. 

Turbulent viscosity is taken into account via the viscous stress tensor 
{\begin{equation}
\mathbf{\Pi} = 2 \Sigma \nu \left( \nabla v - \frac{1}{3}(\nabla \cdot v) {e}  \right),
\end{equation}
where $\nu$ is the kinematic viscosity, $\nabla v$ is a symmetrized velocity gradient tensor and ${e}$ is a unit tensor.}
We parameterize the magnitude of kinematic viscosity $\nu=\alpha_{\rm eff} c_{\rm s} H$, where $c_{\rm s}$ is the sound speed and $H$ is the vertical scale height, using the $\alpha$-prescription of \citet{SS1973} with an effective and adaptive $\alpha$-parameter expressed in the following  form to take the layered disk structure into account \citep{Bae2014}: 

\begin{equation}
\alpha_{\rm eff}=\frac{\Sigma_{\rm a} \alpha_{\rm a} + \Sigma_{\rm d} \alpha_{\rm d}}{\Sigma},
\label{eq:alpha}
\end{equation}
where $\Sigma_{\rm a}$ is the gas surface density of the upper MRI-active 
layer of the disk and $\Sigma_{\rm d}$ is the gas surface density of the lower MRI-dead layer. The total surface density 
of gas is then $\Sigma= \Sigma_{\rm a} + \Sigma_{\rm d}$. {We note that $\Sigma_{\rm d}$ is set equal to zero if $\Sigma \leq \Sigma_{\rm a}$}.
The parameters $\alpha_{\rm a}$
and $\alpha_{\rm d}$ represent the strength of turbulent viscosity in the MRI-active and MRI-dead
layers of the disk. Following \citet{Bae2014} we set $\alpha_{\rm a}=0.01$. Although the actual values
may vary in wide limits, the averaged value is unlikely to significantly exceed this limit \citep[e.g.][]{Yang2018}. 
Indeed,  numerical hydrodynamics simulations of the disk evolution
with a constant $\alpha_{\rm a}=0.1$ yielded disk lifetimes much smaller
than the observationally inferred values of 2--3~Myr \citep{VB2009}. Furthermore, we define $\alpha_{\rm d}$ as 
\begin{equation}
\alpha_{\rm d} = \alpha_{\rm MRI,d} + \alpha_{\rm rd},
\end{equation}
where
\begin{equation}
\alpha _{\rm MRI,d} =
\left\{
\begin{array}{c}
\alpha_{\rm a}, \,\, \mathrm{if} \,\, T_{\rm mp} > T_{\rm crit}   \\
0, \,\,\, \mathrm{otherwise},
\label{MRI_dead}
\end{array}
\right. 
\end{equation}
and $T_{\rm crit}$ is the MRI activation temperature due to thermal effects, such as dust sublimation and/or ionization of alkaline metals and 
$T_{\rm mp}=\mu {\cal P}/(\cal{R}$$ \Sigma)$ is the disk midplane temperature. 
Here, $\mu=2.33$ is the mean molecular weight and $\cal R$ is the universal gas constant. We note that this definition of $T_{\rm mp}$ neglects possible vertical variations in the gas temperature. For a more accurate calculation of the midplane temperature one needs to solve the radiative transfer equations, as was done in the recent modification of the thin-disk limit by \citet{VP2017}.

As was noted in \citet{Bae2014} based on magnetohydrodynamics simulations 
of \citet{Okuzumi2011}, the disk dead layer can have some non-zero residual
viscosity, with $\alpha_{\rm rd}$ as large as $10^{-3}$--$10^{-5}$, due to
hydrodynamic turbulence driven by the Maxwell stress in the
disk active layer. Therefore, we define the residual viscosity as
\begin{equation}
\alpha_{\rm rd} = \min \left( 10^{-5} , \alpha_{\rm a} \frac{\Sigma_{\rm a}}{\Sigma_{\rm d}} \right).
\label{eq:alphaRD}
\end{equation}
This expression takes into account the fact that accretion in the dead zone cannot exceed that
of the active zone if the non-zero $\alpha_{\rm rd}$ is due to turbulence propagating from the
active layer down to the disk midplane.

The cooling and heating rates $\Lambda$ and $\Gamma$ take the disk
cooling and heating due to stellar and background irradiation
into account based on the analytical solution of the radiation transfer
equations in the vertical  direction
\citep[see][for detail]{Dong2016}\footnote{The cooling and heating rates in \citet{Dong2016}
are written for one side of the disk and need to be multiplied by a factor of 2.}:
\begin{equation}
\Lambda=\frac{8\tau_{\rm P} \sigma T_{\rm mp}^4 }{1+2\tau_{\rm P} + 
\frac{3}{2}\tau_{\rm R}\tau_{\rm P}},
\end{equation}
where  $\sigma$ the Stefan-Boltzmann constant, 
$\tau_{\rm R}$  and $\tau_{\rm P}$ are the  Rosseland and Planck
optical depths to the disk midplane, and  $\kappa_P$ 
and $\kappa_R$ (in cm$^{2}$~g$^{-1}$) are the Planck and Rosseland mean opacities taken from \citet{Semenov2003}.

The heating function per surface area of the disk is expressed as
\begin{equation}
\Gamma=\frac{8\tau_{\rm P} \sigma T_{\rm irr}^4 }{1+2\tau_{\rm P} + \frac{3}{2}\tau_{\rm R}\tau_{\rm
P}},
\end{equation}
where $T_{\rm irr}$ is the irradiation temperature at the disk surface 
determined from the stellar and background black-body irradiation as
\begin{equation}
T_{\rm irr}^4=T_{\rm bg}^4+\frac{F_{\rm irr}(r)}{\sigma},
\label{fluxCS}
\end{equation}
where $F_{\rm
irr}(r)$ is the radiation flux (energy per unit time per unit surface
area)  absorbed by the disk surface at radial distance  $r$ from the
central star. The latter quantity is calculated as 
\begin{equation}
F_{\rm irr}(r)= \frac{L_\ast}{4\pi r^2} \cos{\gamma_{\rm irr}},
\label{fluxF}
\end{equation}
where $\gamma_{\rm irr}$ is the incidence angle of radiation arriving at
the disk surface (with respect to the normal) at radial distance $r$. The
incidence angle is calculated using a flaring disk surface as described
in \citet{VB2010}. The stellar luminosity $L_\ast$ is the sum of the
accretion luminosity  $L_{\rm \ast,accr}=(1-\epsilon) G M_\ast \dot{M}/2
R_\ast$ arising from the gravitational energy of accreted gas and the
photospheric luminosity $L_{\rm \ast,ph}$ due to gravitational
compression and deuterium burning in the stellar interior. The stellar
mass $M_\ast$ and accretion rate onto the star $\dot{M}$ are determined
using the amount of gas passing through the inner computational boundary. 
The properties of
the forming protostar ($L_{\rm \ast,ph}$ and radius $R_\ast$) are
calculated using the stellar evolution tracks provided by the \software{STELLAR} code \citep{Yorke2008}. The fraction of
accretion energy absorbed by the star $\epsilon$ is set to 0.05.

\subsection{The inner computational boundary}
\label{subsec:boundary}

The choice of the inner boundary condition in disk formation simulations always presents a certain challenge. 
Ideally, the inner boundary should be placed near the stellar surface, where  the inner disk joins the stellar magnetosphere. 
Such a small inner boundary is however computationally prohibitive even for 2D time-explicit hydrodynamics simulations, as our own, because of strict timestep limitations imposed by the Courant condition \citep{Courant1928} in Keplerian disks. 
This necessitates the use of a sink cell that cuts out the inner disk region, thus moving the inner disk boundary to a larger distance and relaxing the Courant condition. 
The farther the sink-disk interface from the star, the longer timestep the code can manage and the longer integration times (or higher numerical resolution) one can afford. 
This, however, comes at a price --- too large a sink cell may  cut out the disk regions that are dynamically important for the entire disk evolution. In this work, the inner boundary of the computational domain was placed at $r_{sc}=0.4$ au, which allowed us to capture the disk behavior occurring at sub--au scale.
To the best of our knowledge, this is the smallest sink cell used in global collapse simulations of prestellar cores. Moving the inner boundary to several au, as in most collapse simulations, would not allow us to capture the MRI triggering, which plays a key role in the evolution of the inner disk.

Another complication involves the type of the inner boundary.
The inner boundary cannot be of reflecting type, because it must allow for matter to flow to the sink cell and ultimately on the nascent protostar. If the inner boundary allows for matter to flow only in one direction, \ie, from the disk to the sink cell, then any wave-like motions near the inner boundary, such as those triggered by spiral density waves in the disk, would result in a disproportionate flow through the sink-disk interface. As a result, an artificial depression in the gas density near the inner boundary develops in the course of time because of the lack of compensating back flow from the sink to the disk.

\begin{figure}[t]
\begin{centering}
\resizebox{\hsize}{!}{\includegraphics{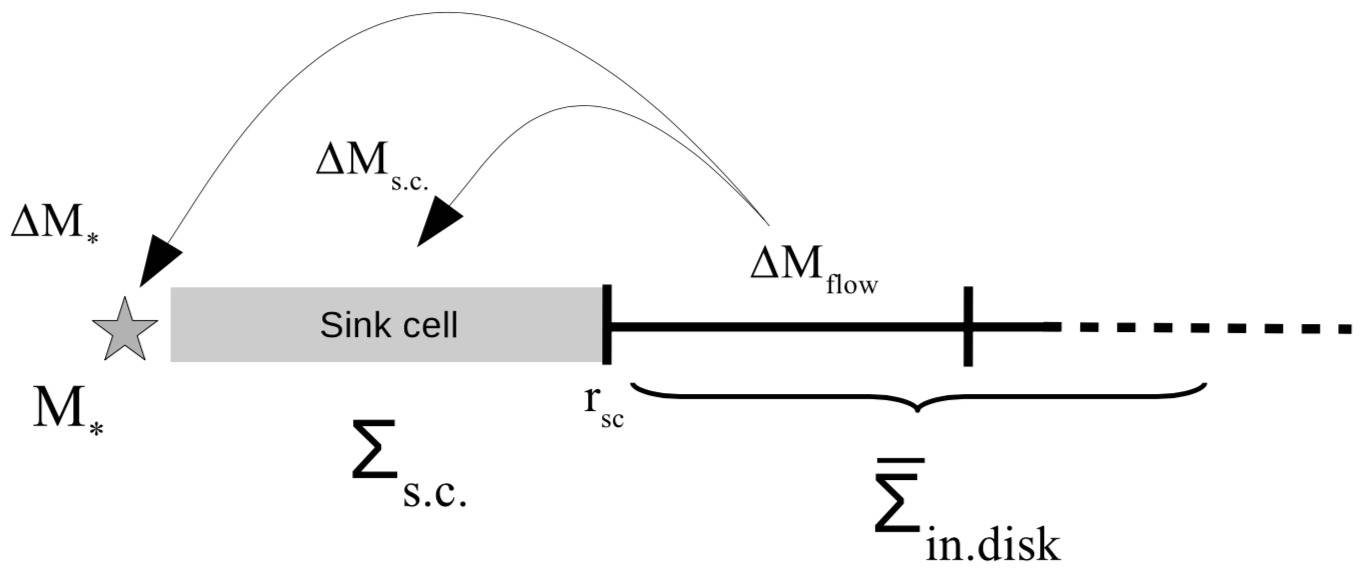}}
\par\end{centering}
\centering{}\protect\protect\protect
\caption{\label{fig:scheme02} 
Schematic illustration of the inner inflow-outflow boundary condition.
The mass of material $\Delta M_{\rm flow}$ that passes to the sink cell
from the active inner disk is further divided into two parts: the mass $\Delta M_\ast$ 
contributing to the  growing central star, and the mass  $\Delta M_{\rm s.c.}$ settling in the sink cell. $r_{sc}$ is the radial position of the sink-disk interface, set equal to 0.4 au.
}
\end{figure}

A solution to this problem is to allow matter to flow freely from the disk to the central sink cell and vice versa. This inflow-outflow boundary condition was presented in \citet{Vorobyov2018} and is schematically illustrated in Figure~\ref{fig:scheme02}. The mass of material that passes to the sink cell through the sink-disk interface is further redistributed
between the central protostar and the sink cell as $\Delta M_{\rm
flow}=\Delta M_\ast + \Delta M_{\rm s.c.}$ according to the following algorithm ({note that $\Delta M_{\rm flow}$ is always positive definite):
\begin{eqnarray}
 \mathrm{if}\,\, \Sigma_{\rm s.c.}^n < \overline{\Sigma}_{\rm in.disk}^n\,\, \mathrm{and}  \,\, v_r(r_{\rm s.c.})<0  \,\, \mathrm{then} \nonumber\\
 \Sigma_{\rm s.c}^{n+1}&=&\Sigma_{\rm s.c.}^n+\Delta M_{\rm s.c.}/S_{\rm s.c.} \nonumber\\
 M_\ast^{n+1}&=&M_\ast^n+\Delta M_\ast \nonumber \\
 \mathrm{if}\,\, \Sigma_{\rm s.c.}^n < \overline{\Sigma}_{\rm in.disk}^n\,\, \mathrm{and}  \,\, v_r(r_{\rm s.c.})\ge 0  \,\, \mathrm{then} \nonumber\\
 \Sigma_{\rm s.c}^{n+1}&=&\Sigma_{\rm s.c.}^n-\Delta M_{\rm flow}/S_{\rm s.c.} \nonumber\\
 M_\ast^{n+1}&=&M_\ast^n \nonumber \\
 \mathrm{if}\,\, \Sigma_{\rm s.c.}^n \ge \overline{\Sigma}_{\rm in.disk}^n\,\, \mathrm{and} \,\, v_r(r_{\rm s.c.})<0 \,\, \mathrm{then} \nonumber\\
 \Sigma_{\rm s.c.}^{n+1}&=& \Sigma_{\rm s.c.}^n \nonumber\\
 M_\ast^{n+1}&=& M_\ast^n + \Delta M_{\rm flow} \nonumber \\
  \mathrm{if}\,\, \Sigma_{\rm s.c.}^n \ge \overline{\Sigma}_{\rm in.disk}^n\,\, \mathrm{and} \,\, v_r(r_{\rm s.c.})\ge0 \,\, \mathrm{then} \nonumber\\
 \Sigma_{\rm s.c.}^{n+1}&=& \Sigma_{\rm s.c.}^n - \Delta M_{\rm flow}/S_{\rm s.c.} \nonumber\\
 M_\ast^{n+1}&=& M_\ast^n. \nonumber
\end{eqnarray}
Here, $\Sigma_{\rm s.c.}$ is the surface density of gas in the
sink cell,  $\overline\Sigma_{\rm in.disk}$ is the averaged surface density
of gas in the inner active disk (the averaging is usually done over
one au immediately adjacent to the sink cell), $S_{\rm s.c.}$ is the surface area of the sink cell, and $v_r(r_{\rm s.c.})$ is the radial component of velocity at the sink-disk interface. We note that $v_r(r_{\rm s.c.})<0$ when the gas flows from the active disk to the sink cell and $v_r(r_{\rm s.c.})>0$ in the opposite case.
}
The superscripts $n$ and $n+1$ denote the current and the updated (next time step) quantities. The exact partition between $\Delta M_\ast$ and $\Delta M_{\rm s.c.}$ is usually set to 95\%:5\%. 
The influence of the $\Delta M_* : \Delta M_{\rm s.c.}$ partition on the disk evolution is studied in \cite{Vorobyov2019}.

The calculated surface densities in the sink cell $\Sigma_{\rm
s.c.}^{n+1}$ are used at the next time step as the inner boundary values for the surface density.
The radial velocity and internal energy at the inner boundary
are determined from the zero gradient condition, while the azimuthal velocity is extrapolated from the active disk to the sink cell assuming a Keplerian rotation.
These inflow-outflow boundary conditions enable a smooth transition of the surface density and angular momentum between the inner active disk and the sink cell, preventing (or greatly reducing) the formation of an artificial drop in the surface density
near the inner boundary. We ensure that our
boundary conditions conserve the total mass budget in the system.
Finally, we note that the outer boundary condition is set to a standard 
free outflow, allowing for material to flow out of the computational
domain, but not allowing  any material to flow in.

\subsection{Initial Conditions and Model Parameters}
\label{subsec:initial}

\begin{table*}
\caption{List of Simulations}
\label{table:sims}
\begin{tabular}{|l|}
\hline
\begin{tabular}{p{2cm}p{1.5cm}p{1.5cm}p{1.5cm}p{1.8cm}}
\hspace{-1cm} Model Name & \hspace{-0.4cm} ${ \rm M_{gas} (M_{\odot}) }$ & \hspace{-0.4cm} ${\rm \beta} (\times 10^{-3})$   & \hspace{-0.2cm} ${\rm T_{crit} (K)}$ & \hspace{-0.4cm} ${\rm \Sigma_{\rm a}}$ (g~cm$^{-2}$) \\
  \end{tabular}\\ \hline
$\kern-\nulldelimiterspace\left.  
\begin{tabular}{l}
$\kern-\nulldelimiterspace\left.
  \begin{tabular}{p{3.5cm}p{1.5cm}p{1.5cm}p{1.5cm}p{1.5cm}}
\simname{model1\_T1300\_S100}$^1$   & 1.152 &  1.360  & 1300 & 100  \\ 
\simname{model1\_T1300\_S10}    & 1.152 &  1.360  & 1300 & 10   \\ 
\simname{model1\_T1500\_S100}   & 1.152 &  1.360  & 1500 & 100  \\
\simname{model1\_T1500\_S10}    & 1.152 &  1.360  & 1500 & 10   \\
  \end{tabular}\right\}$ Magnetically layered disk
\\ 
$\kern-\nulldelimiterspace\left.
  \begin{tabular}{p{3.5cm}p{1.5cm}p{1.5cm}p{1.5cm}p{1.5cm}}
\simname{model1\_const\_alpha}$^{2}$  & 1.152 &  1.360  & -- & --     
  \end{tabular}\right\}$ Fully MRI active disk
\end{tabular}  
\right\}$ Solar mass  
\\ \hline
$\kern-\nulldelimiterspace\left.  
\begin{tabular}{l}
$\kern-\nulldelimiterspace\left.
  \begin{tabular}{p{3.5cm}p{1.5cm}p{1.5cm}p{1.5cm}p{1.5cm}}
\simname{model2\_T1300\_S100}   & 0.346 &  1.355  & 1300 & 100  \\ 
\simname{model2\_T1300\_S10}    & 0.346 &  1.355  & 1300 & 10   \\ 
\simname{model2\_T1500\_S100}   & 0.346 &  1.355  & 1500 & 100  \\
\simname{model2\_T1500\_S10}    & 0.346 &  1.355  & 1500 & 10   \\
  \end{tabular}\right\}$ Magnetically layered disk
\\ 
$\kern-\nulldelimiterspace\left.
  \begin{tabular}{p{3.5cm}p{1.5cm}p{1.5cm}p{1.5cm}p{1.5cm}}
\simname{model2\_const\_alpha}  & 0.346 &  1.355  & -- & --       
  \end{tabular}\right\}$ Fully MRI active disk
\end{tabular}  
\right\}$ Lower mass  
\\
\hline
\end{tabular}\\
    \vspace{1ex}
     \small
      $^1$ Fiducial layered disk model, $^2$ Fiducial fully MRI active disk model
\end{table*}

Our numerical simulations start from the gravitational collapse of a starless cloud core, and continue into the embedded phase of star formation when a star, disk and envelope are formed.
{
The initial surface density and angular velocity of the cloud core are similar to those derived by \cite{Basu1997}, for axially symmetric core collapse
\begin{equation}
   \Sigma = \frac{r_0 \Sigma_0}{\sqrt{r^2+r_0^2}},
 \end{equation}
 \begin{equation}
   \Omega = 2 \Omega_0 {\left( \frac{r_0}{r} \right)}^2  \left[  \sqrt{1+ {\left( \frac{r}{r_0} \right)}^2}  -1 \right]. 
   \end{equation}
Here, $\Sigma_0$ and $\Omega_0$ are the corresponding values at the center of the core and $r_0$ is the radius of the central plateau.     
These parameters depend on the intended initial mass and rotational energy of the core \citep{VB2010}.} 
The simulations are terminated after about 0.5 Myr in the T~Tauri phase, when most of the envelope has accreted onto the central star plus disk system.
We conducted a total of ten disk simulations which are listed in Table~\ref{table:sims}. The prescript \simname{model1} or \simname{model2} corresponds to the total mass of the gas in the initial prestellar cloud core, which is set at 1.152 and 0.346 $M_{\odot}$ respectively. 
With the two parent core masses we aim to demonstrate the dependence of the disk structure and evolution on the final stellar mass.
With a larger core mass \simname{model1} should ultimately form a sun-like star and \simname{model2}, which starts with much smaller cloud core mass, should form a fully convective, sub-solar-mass star. 
All of the initial cloud cores were constructed such that the ratio of the rotational to the gravitational energy, $\beta$, was approximately $0.135\%$. 
This value is consistent with the observations of pre-stellar cores \citep{Caselli2002}.

The MRI driven turbulence is thought to be one of the dominant sources of angular momentum transport in protoplanetary disks \citep*{HGB1995,Turner2014}.
This mechanism acts in a shearing Keplerian disk when the gas in the disk is sufficiently ionized and weakly coupled to the magnetic field.
The midplane temperature in a typical protoplanetary disk is often not high enough to sustain collisional ionization, and thus it forms a magnetically dead zone.
The critical temperature, ${T_{\rm crit}}$, corresponds to the threshold temperature at which the gas in the disk is sufficiently ionized, so that the disk makes an effective transition from a MRI dead state to an MRI active one.
The exact value of ${\rm T_{crit}}$ is difficult to calculate, as it depends on local conditions in the disk such as the distribution of alkali metals, ionization rate, abundance of small grains and threshold ionization fraction for MRI turbulence. 
However, there is a general consensus that for protoplanetary disks the transition occurs abruptly at $\gtrapprox 1000$ K when there is an almost exponential rise in the electron fraction in the gas, primarily due to the ionization of potassium \citep{Umebayashi1983, Gammie1996}.
In accordance with recent investigations \citep[\eg][]{Zhu2010a,Bae2014} we conducted simulations with two values of ${\rm T_{crit}}$, $1300$K and $1500$K, which is denoted in the model name by the number after letter ``\simname{T}''.

Another uncertainty in the model arises due to the thickness of the active layer, which depends on the level of disk ionization.
The cosmic rays are considered to be the dominant source of ionization, penetrating an approximately constant column density of about 100 g~cm$^{-2}$ throughout the disk \citep{UN1981}, which is a canonical value for layered disk models. 
Hard X-rays ($\sim 5$ keV) originating from the star are usually neglected, since 
their penetration depth is typically an order of magnitude smaller \citep{IG1999}.
However, the cosmic rays may not always reach the disk surface unattenuated. 
Similar to the action of solar wind \citep{ST1968}, it is possible that the powerful stellar winds or disk winds of a young star can shield the inner disk from the high energy cosmic rays.
The magnetic field structure of the system can also hinder the propagation of the cosmic rays \citep{Padovani2011}. 
These factors could result in a shielding effect, significantly diminishing active layer thickness.
Thus in our simulations we used two values of ${\rm \Sigma_{\rm a}}$, 100 and 10 g~cm$^{-2}$, representing an unshielded and shielded disk respectively. 
The value of ${\rm \Sigma_{\rm a}}$ is denoted in the model name by the number after letter ``\simname{S}''. 

The first model, \simname{model1\_T1300\_S100} is treated as a representative, fiducial magnetically layered disk model and \simname{model1\_const\_alpha} is treated as fiducial fully MRI active disk model in this paper.
The two models are presented in detail, and all other models are evaluated in comparison with these fiducial models.
Apart from the three parameters --- $M_{\rm gas}$, ${\rm T_{\rm crit}}$ and ${\rm \Sigma_{\rm a}}$ --- all other model parameters are kept identical across the ten simulations. 

The radial and azimuthal resolution of the computational grid was $512 \times 512$ for all simulations, { with logarithmic spacing in the radial direction and uniform spacing in the azimuthal direction. The highest grid resolution in the radial direction was $8.241 \times 10^{-3}$ au at the innermost cell of the disk ($0.4167$ au), while the poorest resolution near the outer boundary of the cloud ($10210$ au for \simname{model1}) was about $202.0$ au.
We assured the sufficiency of the above resolution as well as the numerical convergence of the simulations by comparing global disk properties of the fiducial models at a lower resolution.}
The value of $\alpha_{\rm a}$ and $\alpha_{\rm d}$ in the layered disk models was $10^{-2}$ and $10^{-5}$ respectively, while the fully MRI active disk models were evolved with a single value of $\alpha= 10^{-2}$.
The initial gas temperature of the prestellar cloud core was set to 15K, which was also the temperature of the background radiation.

\section{Results}
\label{sec:results}

\subsection{Dead Zone Formation and Defining Properties}
\label{subsec:properties}
In this section we elaborate on the properties of the gaseous rings formed inside the dead zones in layered disk models. The inner disk structure of the fiducial layered disk model (\simname{model1\_T1300\_S100}) is compared with that of the fiducial fully MRI active disk model (\simname{model1\_const\_alpha}). Both models were evolved from identical initial conditions, and the only difference between them was the viscosity prescription as described in Section \ref{subsec:initial}. 
The classical approach for observational modeling of disks is to consider a Shakura-Sunyaev prescription of viscosity with a constant value of $\alpha$-parameter, which in our case represents a fully MRI active disk.
Although such prescription has been demonstrated to work very well in certain astrophysical objects (\eg, disks in cataclysmic variables, X-ray binaries, see \cite{apia}), a magnetically layered protoplanetary disk is better represented with an adaptive $\alpha$.
The motivation behind this comparison between the two fiducial models was to demonstrate the differences and their implications, especially in the inner disk smaller than about 15 au, which is the region most interesting for planet formation.

\begin{figure*}[t]
\centering
  \vspace{0cm}
\subfloat{
  \includegraphics[width=18cm]{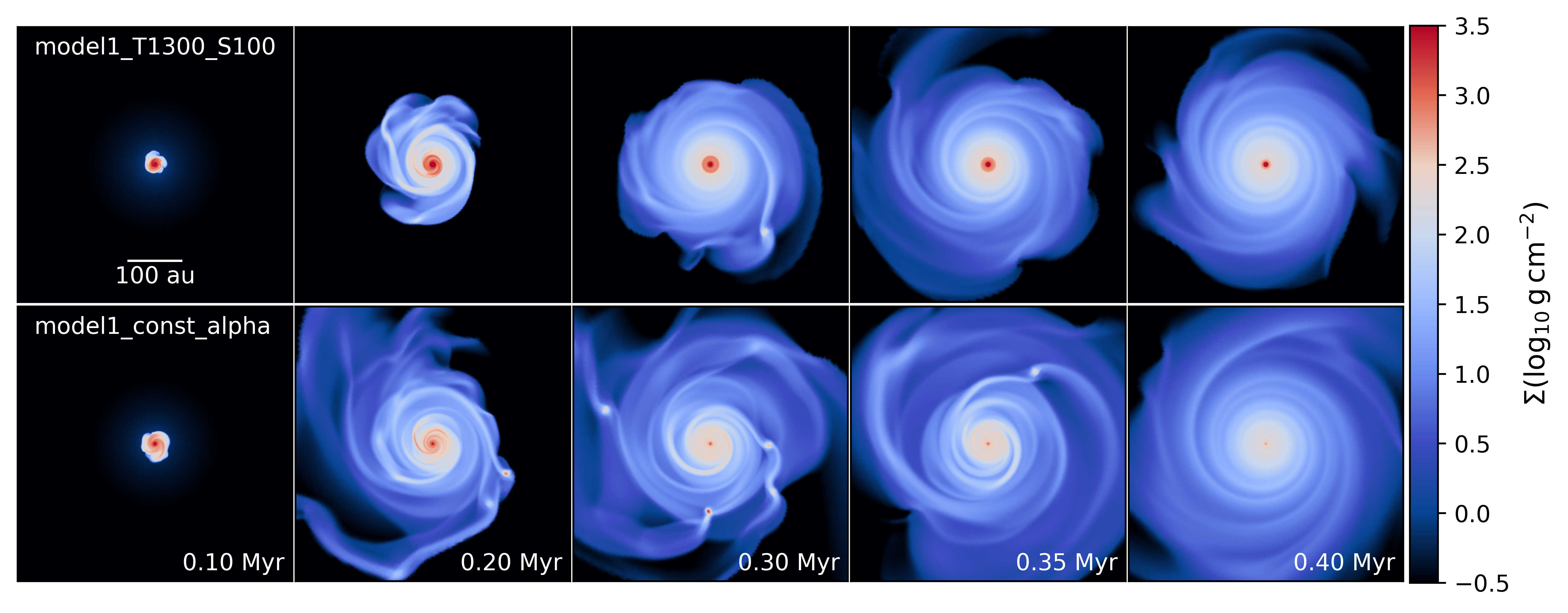}
}
\caption{Evolution of the disk gas surface density distribution for the fiducial layered disk model --- \simname{model1\_T1300\_S100} (top row) --- and fiducial fully MRI active model --- \simname{model1\_const\_alpha} (bottom row) --- over a region of 500 x 500 au. Note that the former consistently shows a higher surface density in the central regions, while the latter shows a larger viscous spread of the disk. }
\label{fig:global}
\end{figure*}

Figure \ref{fig:global} shows the global evolution of the gas surface density distribution in the disk for the two fiducial models over the inner $500\times500$ au region. 
The prestellar cloud core was constructed such that it was gravitationally unstable, and the time was measured from the beginning of the simulations when it began to collapse. 
The disk was evolved up to about 0.5 Myr in both simulations, when it can be considered as an early T Tauri object.
Note that the two models look qualitatively similar over large scale. The initial phases, when the disk is massive, are characterized by gravitational instability and fragmentation. Both disks show irregular spiral structure and formation of clumps in this phase. As the evolution progressed the disks grew and viscously spread, ultimately showing a smoother profile. 
The major difference between the two models is that the central region of the layered disk model showed a consistently higher surface density as compared to the fully active disk. 
The overall size of the disk is smaller in the layered disk model, indicating less viscous spread as compared to the fully MRI active model. 
As a consequence of relatively larger transport of angular momentum to the outer disk region, the fully MRI active disk also shows more clump formation through fragmentation.

\begin{figure*}
\hspace{3.5cm}  Azimuthal profiles    \hspace{2.7cm} \simname{model1\_T1300\_S100}  \hspace{0.8cm}  \simname{ model1\_const\_alpha}\\
\begin{tabular}{c}
\vspace{-0.18cm}\includegraphics[width=0.95\textwidth]{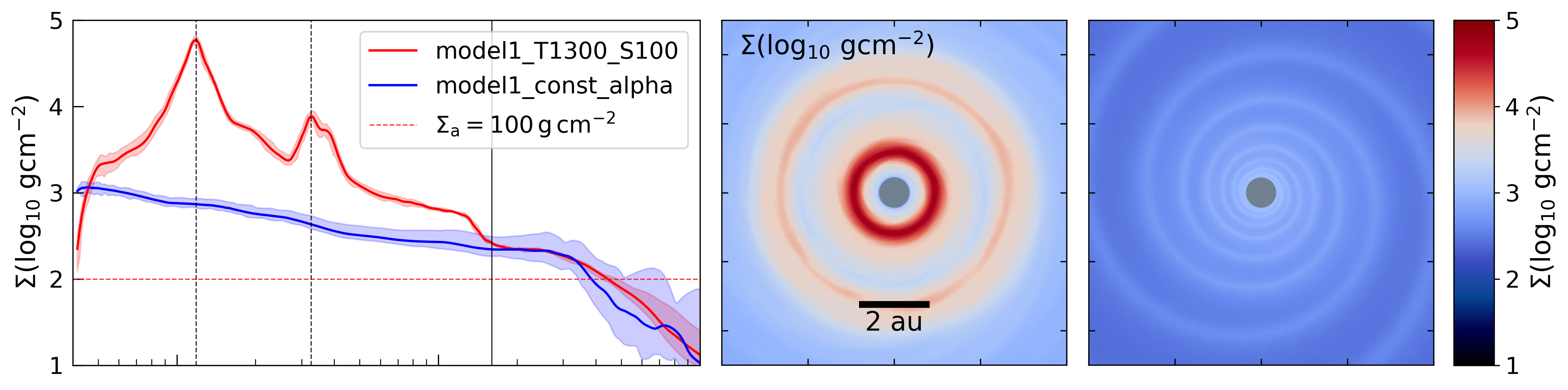} \\ 
\vspace{-0.15cm}\includegraphics[width=0.99\textwidth]{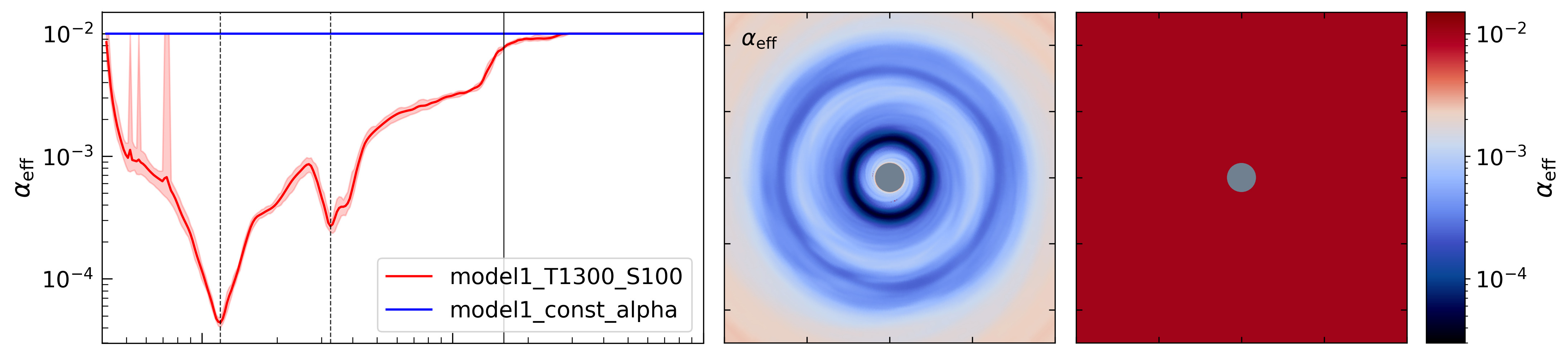} \\ 
\vspace{-0.29cm}\includegraphics[width=\textwidth]{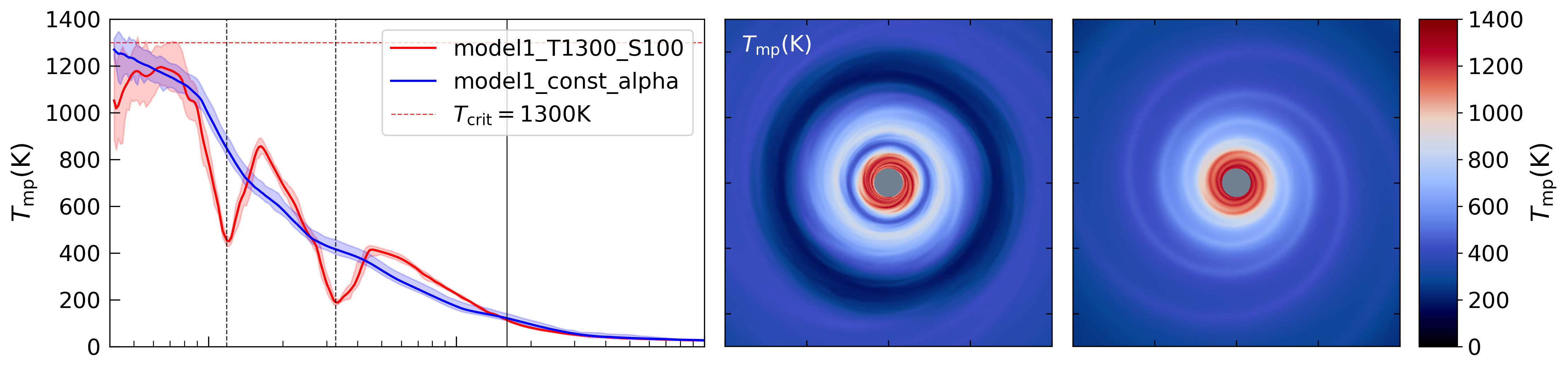} \\
  \includegraphics[width=0.97\textwidth]{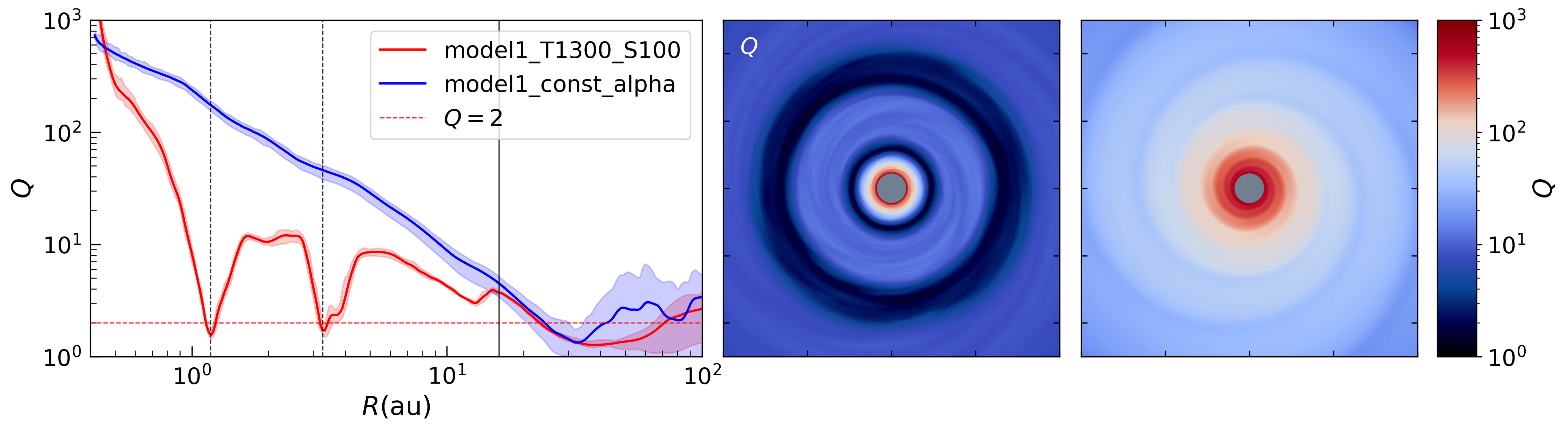} \\
\end{tabular}
\caption{Typical inner disk structure of the fiducial layered disk model is compared against that of the fully MRI active disk model at 0.35 Myr, with respect to quantities --- gas surface density, $\alpha_{\rm eff}$, midplane temperature and Toomre's $Q$-parameter. 
Left: Profiles of the azimuthally averaged parameters, with the shaded area showing the extent between the maximum and the minimum value at the given radius in the inner 100 au region. The two vertical dashed lines mark the location of the two rings, while the solid line marks the outer extent of the dead zone. Right: The distribution of the same parameters is plotted in the inner $10 \times 10$ au box.}
\label{fig:inner1}
\end{figure*}

In Figure \ref{fig:inner1} we focus on the inner region of the fiducial disks at 0.35 Myr. 
This time corresponds to the fourth column in Figure \ref{fig:global}, when the disk structures are typical for both simulations. 
At this point, the disks had started to settle down, became increasingly symmetric and showed less flocculent spiral structure. 
Figure \ref{fig:inner1} compares a snapshot of four quantities of the fiducial disks --- total surface density, ${\rm \alpha_{eff}}$ as given by Equation \ref{eq:alpha}, midplane temperature, and Toomre's $Q$-parameter (calculated as $Q=c_{\rm s} \Omega/  \pi G \Sigma$). 
In each row, the two images on the right hand side show the distribution of the variables in the inner $10\times10$ au box. On the left hand side in each row the azimuthally averaged profiles of the same variables are plotted for comparison in 100 au radius. 
The shaded region shows the extent between the maximum and minimum values at a given radius.
Note that the abscissa in the azimuthal profiles are in logarithmic units.

We define the dead zone in the layered disk model as the region where ${\rm \alpha_{eff}}$ falls below 80\% of the fully MRI active value of ${\rm \alpha_{a}}= 0.01$.
The physical motivation behind this definition was that below this threshold of ${\rm \alpha_{eff}}$, the gas surface density in all of the layered models started to diverge from a fully MRI active disk, as can also be observed in Figure \ref{fig:inner1}.
It was possible to consider the dead zone in terms of the surface density, as dead zone exists only when $\Sigma \geq \Sigma_{\rm a}$.
However, we will show that this was always an overestimate, and the above definition of dead zone based on the viscosity prescription (${\rm \alpha_{eff}}$) reflected the status of the inner disk more accurately.
The vertical black line in the azimuthal profiles in Figure \ref{fig:inner1} marks the outer extent of the dead zone at this time, which was at about 16 au.
The azimuthal profiles of the quantities from the two simulations in Figure~\ref{fig:inner1} tend to converge beyond this point and within the dead zone they differed significantly.

The most compelling difference between the fiducial models was the formation of concentric, axisymmetric, gaseous ``rings'' within the dead zone in the layered disk model.
As seen in azimuthal plots as well as the spatial distributions, the rings offer a remarkable contrast in the local conditions as compared to the fully magnetically active disk.
These rings are characterized by a much higher surface density and a lower effective viscosity, indicated by a low value of ${\rm \alpha_{eff}}$. 
The two vertical lines at about 1.2 and 3.2 au in the azimuthal plots indicate the two maxima in the surface density and hence the position of the two rings. 
For the most prominent ring, the surface density was about two orders of magnitude higher and the local ${\rm \alpha_{eff}}$ was about two orders of magnitude lower as compared to the corresponding values in the fully active disk.
The outer ring, although clearly distinguishable, was about an order of magnitude less pronounced, in terms of both $\Sigma$ and $\alpha_{\rm eff}$, as compared to the inner ring.
The midplane temperature at the locations of the rings was lower, as compared to a fully active disk, because of the lower viscous heating in these regions. 

Toomre's $Q$-parameter compares the destabilizing effects of self--gravity in the disk against the stabilizing effects of pressure and rotation. 
This parameter is expected to increase as we get closer to the star due to the increased rotational velocity and rising temperature, as we see in the fully MRI active model. 
However, due to the accumulation of gas as well as the lower midplane temperature, the $Q$-parameter was low ($\approx 2$) in the rings which indicates that this region was marginally gravitationally unstable.

\subsection{Dynamical Behavior of the Rings}
\label{subsec:dynamical}
\begin{figure*}
\hspace{3.3cm} \simname{model1\_T1300\_S100} \hspace{4.6cm} \simname{model1\_const\_alpha}  \\
\begin{tabular}{c}
\hspace{-0.56cm}  \includegraphics[width=0.95\textwidth]{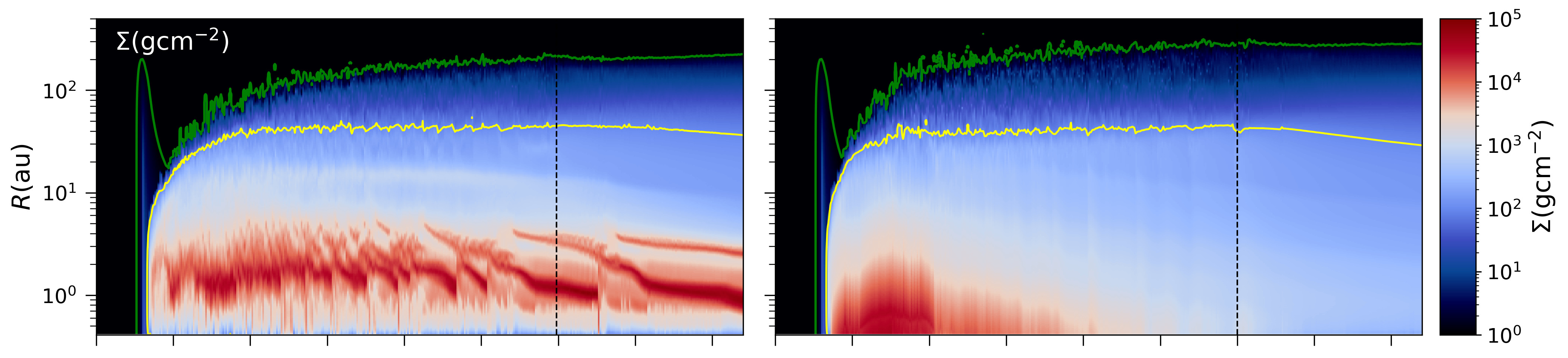} \\ \hspace{-0.55cm}  \includegraphics[width=0.955\textwidth]{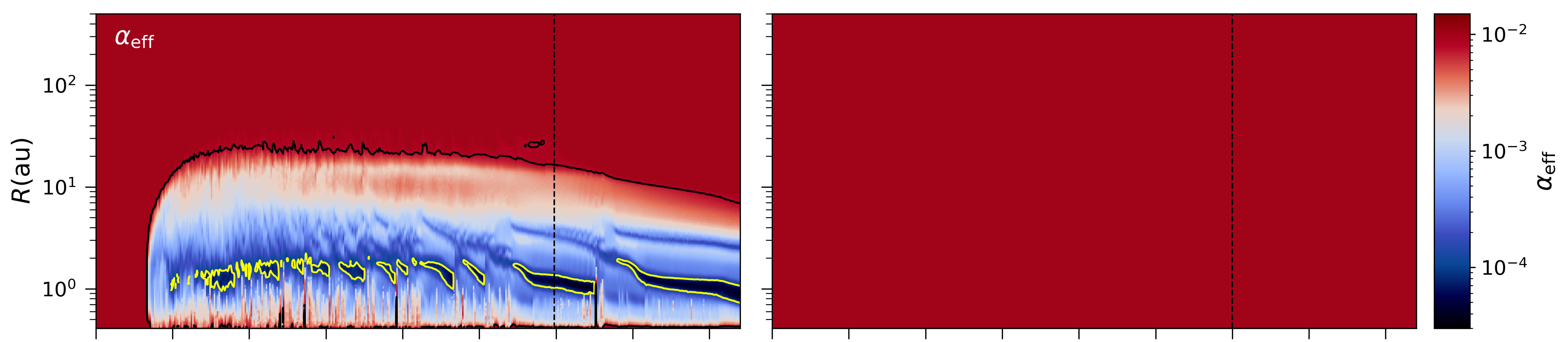} \\
 \hspace{-0.55cm} \includegraphics[width=0.95\textwidth]{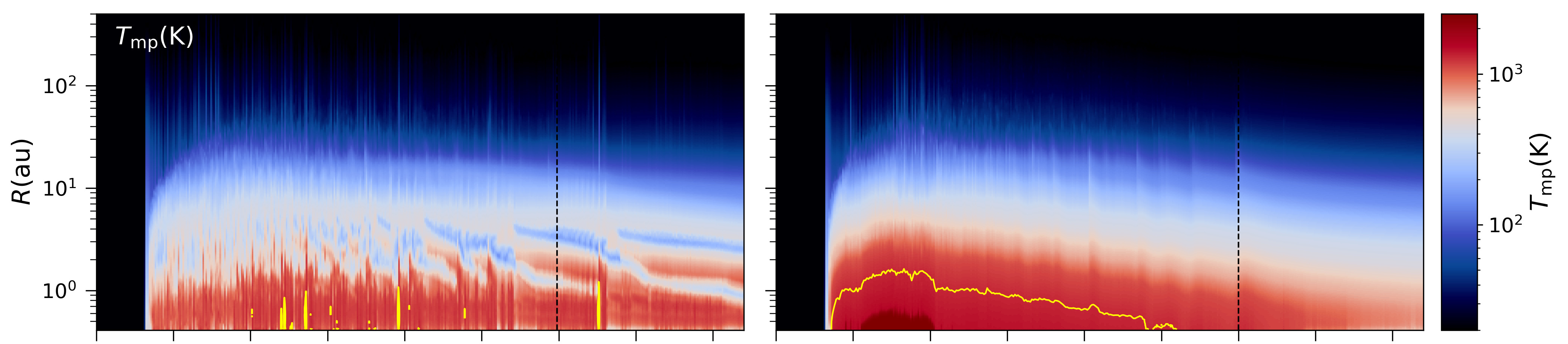} \\  
\hspace{-0.64cm} \includegraphics[width=0.95\textwidth]{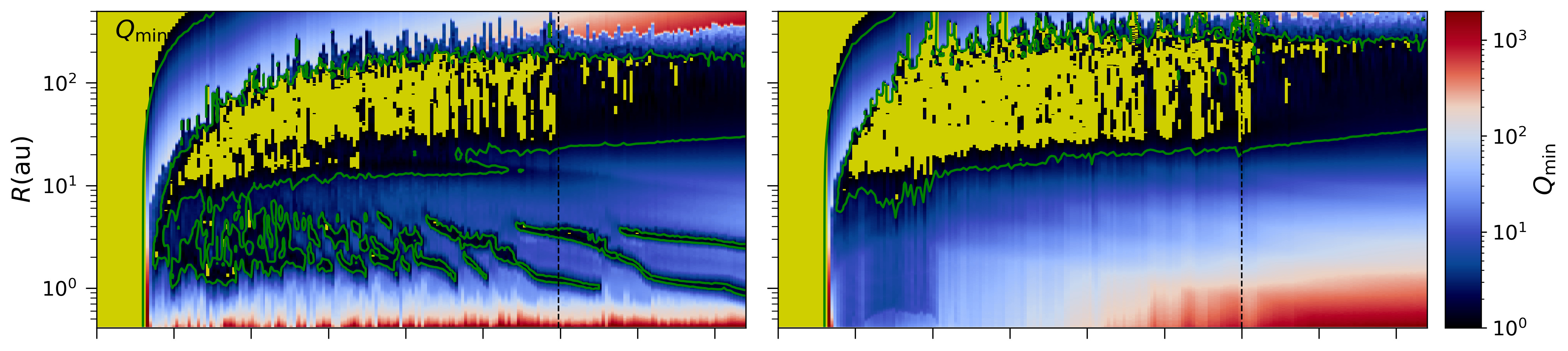} \\
\vspace{-0.5cm}
\hspace{0.05cm} \includegraphics[width=0.99\textwidth]{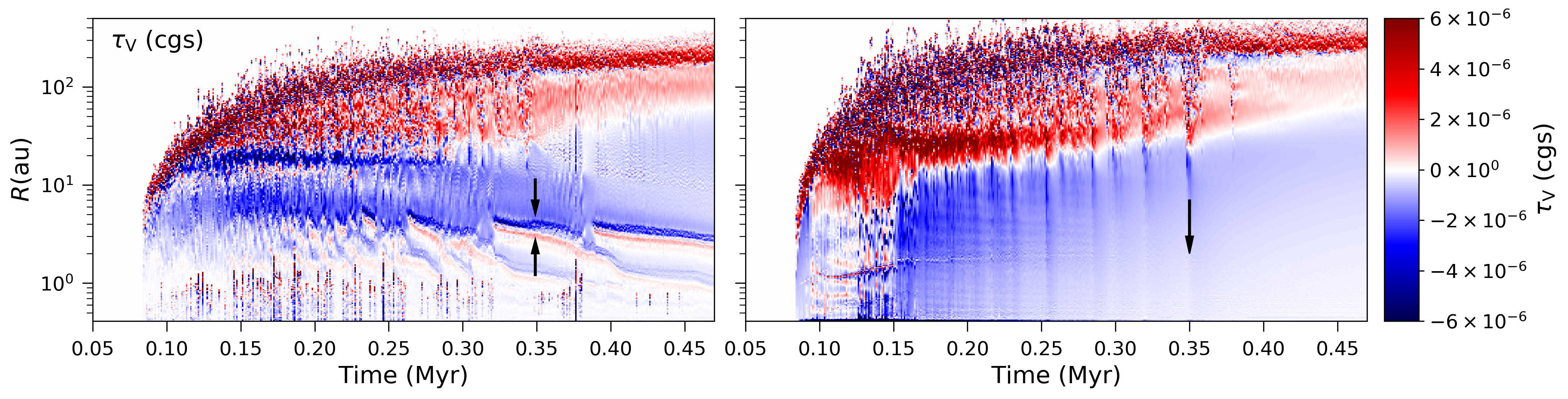} \\
\end{tabular}
\caption{Spacetime plots for the two fiducial models with the rows depicting the evolution of azimuthally averaged quantities $\Sigma$, $\alpha_{\rm eff}$ and $T_{\rm mp}$, the minimum in $Q$-parameter {and viscous torque}. 
The green and yellow curves in $\Sigma$ show 1 and 100 ${\rm g~cm^{-2}}$ contours respectively.
The black contour in $\alpha_{\rm eff}$ shows the extent of the dead zone as defined in Section \ref{subsec:properties}, while the yellow lines shows $10^{-4}$ contour. 
The yellow lines in the midplane temperature plots show $T_{\rm crit}=1300$ K contours. The yellow shaded area in $Q$-parameter plot shows gravitationally unstable region where $Q\leq1$, {while the green contours show $Q=2$ level.}
The layered disk model (first column) shows qualitatively different behavior with the formation of long lived rings, which can be clearly seen as high surface density and low viscosity bands. The vertical dashed lines show the time corresponding to the static images in Figure \ref{fig:inner1}, {while the arrows in the last row show the direction of viscous torques, which act towards building up of the rings in the layered disk. }
}
\label{fig:spacetime1}
\end{figure*}

Although the properties of the inner disk region in Section \ref{subsec:properties} are typical for the layered disk model, the rings themselves are highly dynamical and constantly evolving structures.
In order to demonstrate their behavior across time, we plot spacetime diagrams in Figure \ref{fig:spacetime1}. This figure compares the azimuthally averaged profiles of quantities $\Sigma$, ${\rm \alpha_{eff}}$ and T, the minimum in Toomre's $Q$-parameter, as well as the {viscous torque ($\tau_\nu$)}. 
Each row shows the evolution of the variables for fiducial layered disk model on the left and fully MRI active disk model on the right.
Note that the ordinate (R) is in logarithmic units. 
The azimuthally averaged profiles were printed every 1000 yr in our numerical code, which determines the temporal resolution in this plot.
The time spacing of $Q$-parameter was 2000 yr, since the minimum value was calculated using two-dimensional output files.
We found that this resolution in time was sufficient for analyzing the dynamical behavior of the disk.
The vertical dashed line at 0.35 Myr denotes the snapshot in time corresponding to Figure \ref{fig:inner1}. 
In accordance with the earlier discussion, the region outside the dead zone was similar for both fiducial models throughout the evolution, while the inner region appeared markedly different.  

Consider the first row in Figure \ref{fig:spacetime1} depicting the evolution of surface density. The green contour plotted at $\Sigma=1$ g~cm$^{-2}$ shows the approximate extent of the disk, and the yellow contour is plotted at ${\rm \Sigma_{a}=100}$ g~cm$^{-2}$. 
With the fully MRI active disk, the surface density approximately monotonically decreases with radius, while also decreasing across time as the disk evolves.
Some variations in the surface density at larger distances and earlier times are the result of clump formation through disk fragmentation, in both simulations.
In the layered disk model, enhancements in surface density in the region between 0.6 and 5 au started occurring as soon as the disk was formed at 0.08 Myr, while axisymmetric structures resembling rings started appearing after about 0.2 Myr.
The rings can be easily identified as the diagonal high surface density bands.
As the time evolved, the rings become denser and they also migrated closer to the star. 
Multiple rings could form at a given radius, however, the innermost ring was typically the most prominent. 
After the inward migration, we can observe that a ring terminated with a sudden discontinuity. 
We found that at these points in time, MRI was triggered and as a result the ring became unstable.
The material in the ring was accreted onto the star and the resulting eruption qualitatively resembled an FUor outburst.
As mentioned earlier, in this paper we will concentrate on the structure and evolution of the inner disk, and the detailed analysis of the outbursts will be conducted in a subsequent paper.

The second row of Figure \ref{fig:spacetime1} compares the effective $\alpha$-parameter, which has a constant value of 0.01 for the fully MRI active disk.
The black contour shows the extent of the dead zone in the layered disk model, defined as the region where the $\alpha_{\rm eff}$ falls below 80\% of this maximum value.  
The yellow contours depict a level of $10^{-4}$. Thus within the vicinity of the rings, $\alpha_{\rm eff}$ is less than one percent of its MRI active value.
This low viscosity region coincides with the rings observed in the gas surface density. 
Note that for the layered disk model, at early times the dead zone was consistent with the ${\rm \Sigma_{a}=100}$ g~cm$^{-2}$ contour. As the disk evolved the gas surface density became exceedingly flatter, and the dead zone was contained within a much smaller radius.
The third row compares the azimuthally averaged midplane temperature. For the fully MRI active disk the temperature distribution mirrored the surface density, which was approximately monotonically decreasing with the radius. The disk was initially hot and then cooled as it evolved.
In contrast, the innermost regions of the layered disk model ($< 1$ au) remained at approximately constant temperature throughout its 0.5 Myr evolution. 
The temperature near the rings was lower than the surroundings because of the decreased viscous heating. The yellow contour shows $T_{\rm crit}=1300 K$. 
Notice that the midplane temperature in the inner disk periodically spiked above this value for a short period in the layered disk model. 
The increases in the inner disk temperature coincide with the disappearance of the rings and were thus associated with the MRI outbursts.

The {fourth row of Figure \ref{fig:spacetime1}} compares the minimum in the Toomre's $Q$-parameter for the two fiducial simulations, which indicates if the disk is gravitationally unstable at a given radius.
The yellow shaded region shows the region where $Q<1$, {while the green contours show the $Q=2$ levels.} 
In both simulations, the initial gravitational collapse phase of a prestellar core can be seen before the disk formation at about 0.08 Myr as gravitationally unstable at all radii.
As expected, disks in both simulations were gravitationally unstable at large distances ($\gtrapprox 10$ au) and {early times ($\lessapprox 0.35$ Myr)}, which points to the formation of clumps via disk fragmentation.
{In the vicinity of the rings, the yellow pigments suggest that the rings showed brief instances of GI fragmentation, although not all such occurrences were captured due to the limited time resolution. 
In this region the $Q$-parameter remained consistently low due to the large total surface density as well as a low temperature.
We elaborate on the role of GI and associated spirals on the evolution of the rings later in this section.
}

{The last row of Figure \ref{fig:spacetime1} shows the viscous torque for the two fiducial models, calculated as $\tau_{\nu}(r,\phi) = r(\nabla \cdot {\Pi})_\phi S(r,\phi)$, where $\nabla \cdot {\Pi}$ is the viscous force per unit area  (equation~\ref{eq:momentum}), $r$ is the lever arm, and $S(r,\phi)$ is the surface area occupied by a cell with coordinates $(r,\phi)$ (see \cite{VB2009} for full expression).   
The outer regions are noisy due to the action of clumps and spirals, especially in the early times.
Neglecting the initial about 0.15 Myr, the general direction of the viscous torques in the inner disk ($\lessapprox 10$ au) is highlighted by arrows.
The fully MRI active disk shows a smooth negative viscous torque in this region, indicating a steady inward mass transfer.
The evolution of $\tau_{\nu}$ in the fiducial layered disk model is again much different; the viscous torques change sign across the rings, and act towards building up the rings from both radial directions. 
The action of viscous torques can explain how the rings form in the inner disk. 
The outer edge of the dead zone usually undergoes a smooth and gradual transition.
As the surface density decreases in the radial direction, the shift from dead to fully active region occurs over several tens of au.
Thus the mass does not accumulate near the outer boundary of the dead zone. 
The inner boundary of the dead zone on the other hand undergoes a sharp transition when the temperature exceeds $T_{\rm crit}$ (1300 K for the fiducial model).
The viscous torque has a component proportional to the negative of the gradient of kinematic viscosity, so the material is accelerated outward at the inner boundary.
In the fiducial layered disk model, the viscosity gradient was relatively large in the vicinity of the rings, as can be inferred from the azimuthal plots of $\alpha_{\rm eff}$ in Figure \ref{fig:inner1}.
Thus when the gas reached the inner regions of the dead zone, the viscous torques worked in the direction of reinforcing the pileup of material.
The resulting positive feedback between viscous torques and gradients of kinematic viscosity (due to the increased surface density) generated the stable ring structures.
The viscous torque is proportional to the distance from the axis of rotation, hence its magnitude decreased in general as the rings moved inward.
}

\begin{figure*}
\hspace{3.5cm}  Azimuthal profiles    \hspace{3.5cm} No GI  \hspace{2.8cm}  With GI\\
\begin{tabular}{c}
\vspace{-0.18cm}\includegraphics[width=\textwidth]{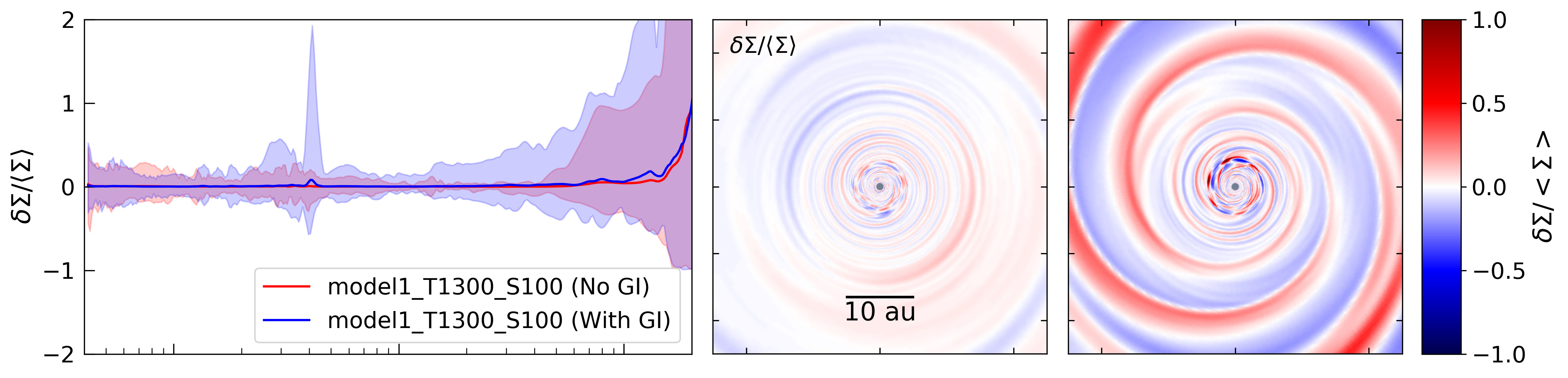} \\ 
\hspace{-0.44cm}\includegraphics[width=\textwidth]{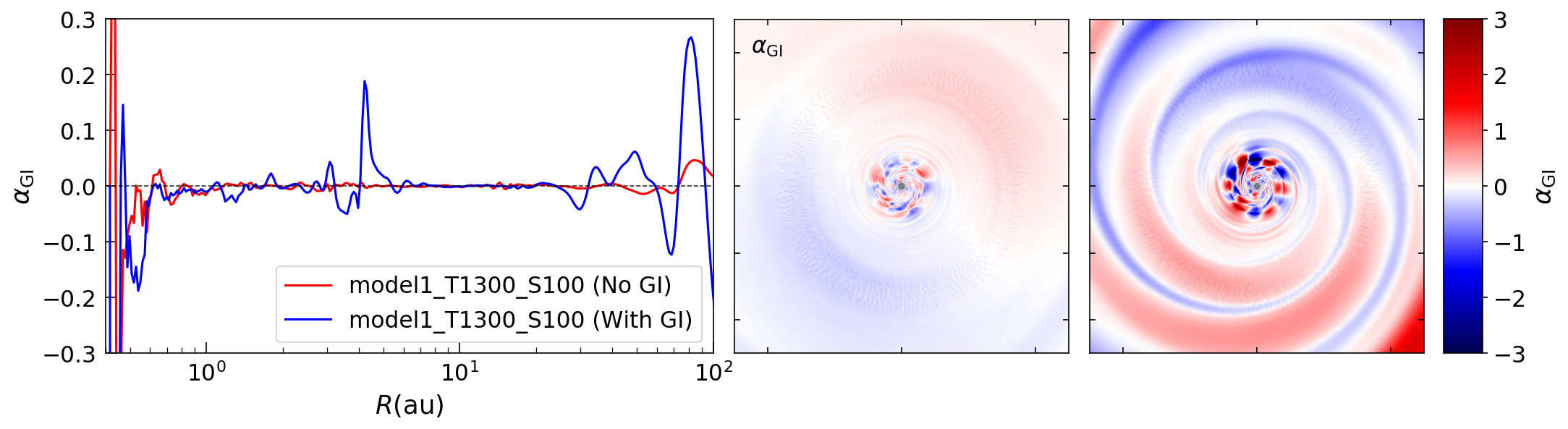} \\ 
\end{tabular}
\caption{The fractional amplitude and $\alpha_{\rm GI}$--parameter of the fiducial layered disk model is compared when GI fragmentation (with $Q<1$) was absent (0.350 Myr) against when it was present (0.326 Myr) in the rings. 
Left: Profiles of the azimuthally averaged fractional amplitudes, with the shaded area showing the extent between the maximum and the minimum value at the given radius in the inner 100 au region. 
Right: The distribution of the same parameter is plotted in the inner $50 \times 50$ au box.}
\label{fig:v2}
\end{figure*}

\begin{figure}
\centering
\includegraphics[width=0.5\textwidth]{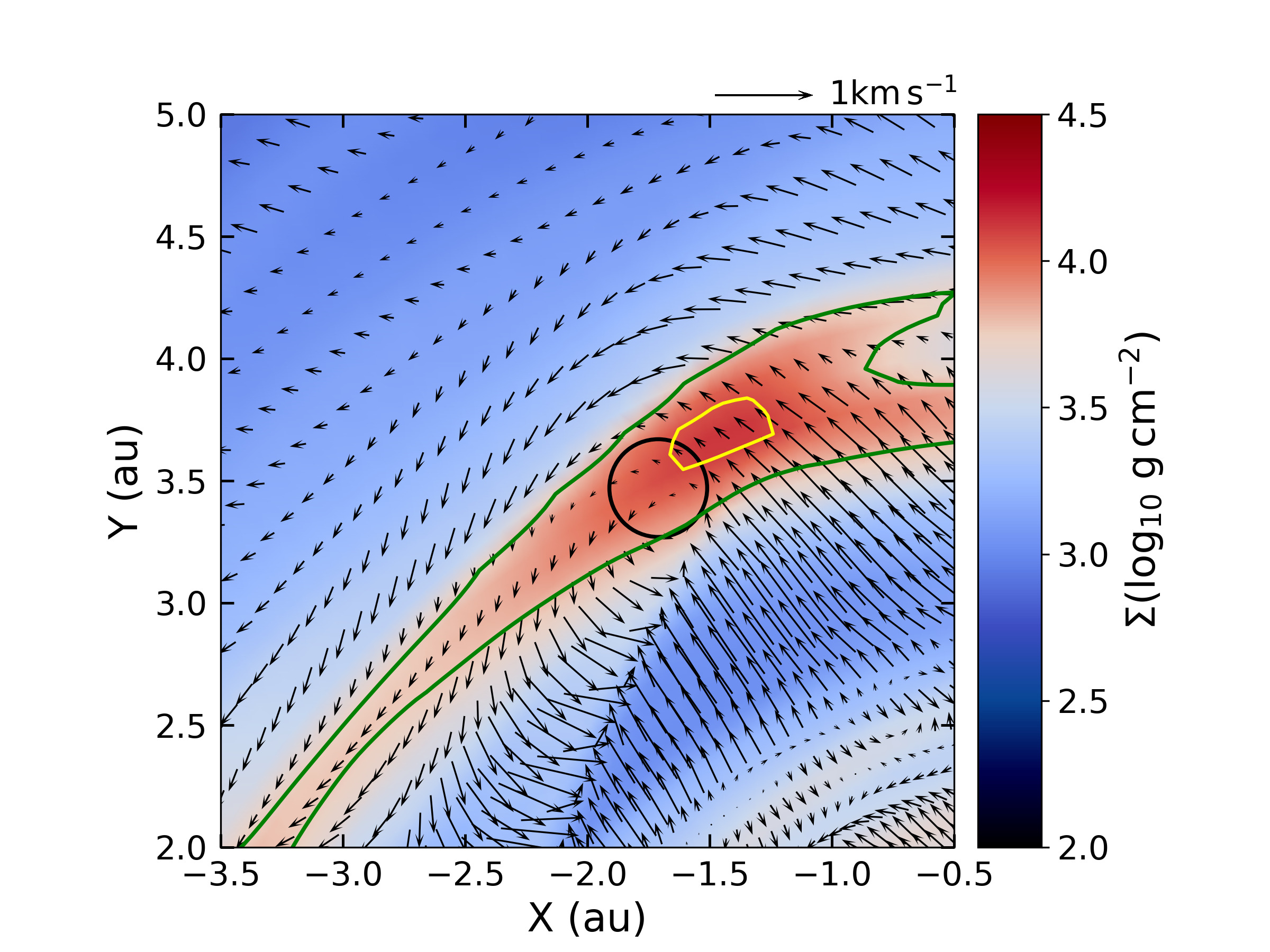}
\caption{The velocity field superimposed on the gas surface density distribution, showing formation of a vortex near the outer ring. {The black circle marks the region of the vortex, while the green and yellow contour lines show the onset of GI with $Q=2$ and $1$ levels respectively.}}
\label{fig:velocity}
\end{figure}

{The interplay between the magnetically layered structure in the form of rings and gravitational instabilities showed interesting, time--dependent manifestations over a large extent of the disk.
Due to the action of viscous torques the rings accumulated enough mass to show occasional gravitational fragmentation.
These fragments could not survive for long due to strong Keplerian shear at these radii, as well as their proximity to other overdense regions of the ring. 
However, this threshold for GI fragmentation ($Q=1$) captures only an extreme case of the non-linear outcome of axisymmetric instability. 
A disk which is locally stable according to this criterion may still generate non-axisymmetric, large-scale spiral waves \citep{Hohl1971,PPS1997}.
Thus the onset of GI may occur at a threshold as high as $Q\approx2$ and associated spiral waves can extend over a large fraction of the disk.
When the disk is massive 
the evolution of GI may show episodes of intense spiral activity followed by quiescent phases \citep{LR2005}.
Similar behavior was observed in our simulations; spiral density waves emanated from periodic asymmetries developed in the rings and they were particularly strong when GI fragmentation occurred.
We demonstrate this in Figure \ref{fig:v2}, which compares the disk properties when gravitational fragmentation was absent in the rings (snapshot at 0.350 Myr, corresponding to Figure \ref{fig:inner1}) to when the rings showed GI fragmentation with $Q<1$ (snapshot at 0.326 Myr, corresponding to Figure \ref{fig:velocity}).
The first row shows the fractional amplitudes, $\delta \Sigma/ \langle\Sigma \rangle$, where the denominator is averaged in the azimuthal direction.
The spiral density waves with low azimuthal mode number (usually $m=2$) originating at locally unstable region extended across the entire dead zone, while such waves were over an order of magnitude weaker at other times when the rings were more axially symmetric.
Note that the fractional amplitude map for the fiducial fully MRI active disk showed spirals as well, which can be inferred from the first row of Figure \ref{fig:inner1}.
These spirals did not show a time-dependent behavior except their progressive weakening as the disk evolved (see \cite{VB2009} for more details).

The large--scale spiral waves can contribute towards angular momentum transport across the dead zone, and thus enable inward migration of the rings.
The second row of Figure \ref{fig:v2} quantifies the angular momentum transport due to GI via gravitational $\alpha$--parameter calculated as a ratio of gravitational stress $G_{xy}$ and vertically averaged pressure $P$
\begin{equation}
\alpha_{\rm GI} = \frac{G_{xy}}{ P}= \frac{1}{4 \pi G P} \frac{\partial \Phi}{\partial x} \frac{\partial \Phi}{\partial y},
\end{equation}
where $\Phi$ is gravitational potential. 
The azimuthally averaged $\alpha_{\rm GI}$ can be added to $\alpha_{\rm eff}$ to give an estimate of total viscosity (and thus the angular momentum transport) at a given radius due to both GI and MRI processes. 
When azimuthal asymmetries were absent in the rings, the inner disk showed consistently low values of $\alpha_{\rm GI}\approx 0$ (the fiducial fully MRI active disk showed similar low values).
In presence of asymmetries with GI fragmentation, $\alpha_{\rm GI}$ was much stronger, reaching about $0.2$ due to strong spiral density waves.
This value is over an order of magnitude larger than the largest possible value of $\alpha_{\rm eff}=\alpha_{\rm a}=0.01$ in the layered disk.
Thus these GI spiral waves were capable of redistributing a significant amount of angular momentum and gravitational energy across the dead zone.
Note that even when the $Q$--criterion is fulfilled locally, the disk needs to cool efficiently over dynamical timescale to form GI fragments \citep{Rafikov2005}. As rings migrated inward, they may be more stable against fragmentation because of inefficient cooling and thus showing a slower rate of migration. 

}

{An additional} aspect of dynamical behavior of the inner disk was occasional formation of vortices in the vicinity of the rings.
Figure \ref{fig:velocity} depicts an instance of a vortex formed near the outer, secondary ring at 0.326 Myr.
The velocity field is plotted with respect to the local Keplerian velocity, and the vortex can be seen in the region marked with the black circle. 
Vortex formation can be triggered by Rossby wave instability which can develop at pressure bumps near a sharp viscosity transition in a protoplanetary disk \citep{Lovelace1999}. 
Such vortices are capable of collecting dust and thus aid formation of planetesimals if they are long-lived.
We found that the vortices formed near azimuthal asymmetries in the gas surface density in a ring, and were usually in proximity with the gravitationally unstable regions occurring in the outermost parts of a ring structure.
{The vortices were always cyclonic} (\ie, rotating in the same direction as the disk) and hence naturally unstable in a Keplerian disk {due to strong shear \citep[\eg][]{GL1999}.
\cite{RV2017} showed that the self--gravitating vortices are subject to strong torques and hence are prone to a significantly faster decay as compared to non self--gravitating ones.
We confirmed the rapid decay of the vortices by restarting parts of the simulation and obtaining a high temporal output of 20 yr.
The vortices completely disappeared within this short time of disk evolution, however, note that their dissipation at 4 au should proceed quicker, over the local dynamical time scale.
Their proximity to GI fragmentation suggests that the instability formed spiral arms and the resulting waves in turn generated local, short--lived vortices.
One caveat of our simulations is that the vortices were relatively small and hence only marginally resolved -- since the grid resolution at 4 au was about 0.08 au in the radial direction. 
However, the vortices were cyclonic and formed near locally unstable regions, both of these factors contribute towards a rapid decay.
Thus additional grid resolution should not change the overall results of the simulations. 
}

In order to study the rings formed in the layered disk model quantitatively, we define a ring to be the structure corresponding to the location of maximum surface density at a given time. 
From the spacetime diagrams in Figure \ref{fig:spacetime1} we can see that a ring typically started its evolution when there was another more prominent ring closer to the star. 
This phase seemed to be about half of the total lifetime of a ring. 
Thus with the above definition we cannot track the entire evolution of a ring, however, we should obtain estimates which are accurate up to a factor of two.  
In Figure \ref{fig:ringquant} we consider the evolution of the fiducial layered disk model between 0.30--0.46 Myr when the rings are fairly stable in time.
The first row shows the location of the ring, where the inward drift can be clearly noticed, as well as large discontinuities occurring at about 0.305 and 0.375 Myr. 
These discontinuities are associated with the instability of the most prominent ring, and correspond to FUor--like outbursts. 
The vertical dashed lines mark these accretion events. 
Neglecting the first 0.2 Myr when the ring structures were highly asymmetric and transient, we calculated the average rate of migration of the rings to be about $-25$ au/Myr (depicted by the diagonal red, dashed line), 
while the average duration between the MRI outbursts was about $38\,000$~yr. 

The next two rows of Figure \ref{fig:ringquant} show the surface density and $\alpha_{\rm eff}$ at the location of the rings. 
As a ring formed a bottleneck for mass transport within the dead zone, the incoming material from the outer disk was accumulated in its vicinity. 
Thus as the ring evolved, its surface density increased almost monotonically. 
The residual viscosity in the dead zone, which is induced by the action of the active surface layers, is parametrized via $\alpha_{\rm d}$ (Equation \ref{eq:alphaRD}). 
Thus as $\Sigma_{\rm d}$ increased, the $\alpha_{\rm eff}$ decreased.
The local value of $\alpha_{\rm eff}$ could jump over two orders of magnitude during the MRI instability.
Note that we did not catch the actual accretion event occurring at 0.305 Myr due to the sparse rate of output (every 1000 yr), but did so for the outburst at 0.375 Myr.
The fourth row of Figure \ref{fig:ringquant} shows the evolution of the midplane temperature at the ring location. 
The temperature exceeded $T_{\rm crit}=1300$ K at the discontinuity at 0.375 Myr, indicating that the instability in the rings was indeed caused by MRI activation.

The last two rows of Figure \ref{fig:ringquant} show the kinematic viscosity ($\nu_{\rm kin}=\alpha {c_s}^2 / \Omega_k$) and the viscous timescale ($T_{\rm visc}=r^2/\nu_{\rm kin}$) at the ring location. 
As the ring evolved, the kinematic viscosity at its location decreased due to the decrease in $\alpha_{\rm eff}$, indicating a narrower bottleneck for mass transport. 
As expected, the $\nu_{\rm kin}$ briefly increased to a large value, and the bottleneck in mass transport disappeared at the accretion events.
The viscous timescale of the rings stayed approximately constant at about 1 Myr.
{This implies that if left alone, the rings themselves would evolve and dissipate over this timescale.
We hypothesize that the inward migration of a ring was primarily facilitated by large--scale spiral waves induced by GI, resulting in a rate of migration faster than the viscous timescale.
The lifetime of a typical ring was much shorter than the viscous timescale because MRI was always triggered before the rings could show viscous dissipation.} 
When MRI was triggered the viscous timescale at the ring location decreased significantly, to about 1000 yr, for the outburst near 0.375 Myr. 
The material was accreted within this very short time as compared to the average viscous timescale of the rings. 
We found that the duration of the luminosity outburst (luminosity was sampled at a higher frequently) was in agreement with this minimum value of the viscous timescale.
{
Notice that in Figure \ref{fig:spacetime1} as the simulation progressed the inward migration time of successive rings increased.
We found that the asymmetries in the rings and GI fragmentation, both of which generated spiral waves, could be induced by strong perturbations in the inner disk -- \eg, an infalling clump or an MRI outburst (described below).
As the simulation evolved, the frequency of the clumps decreased in general (see Figure 4, $Q$--parameter), which should result in less GI spirals and the observed slower migration rate of the rings.

\begin{figure}
\centering
\subfloat{
  \includegraphics[width=0.46\textwidth]{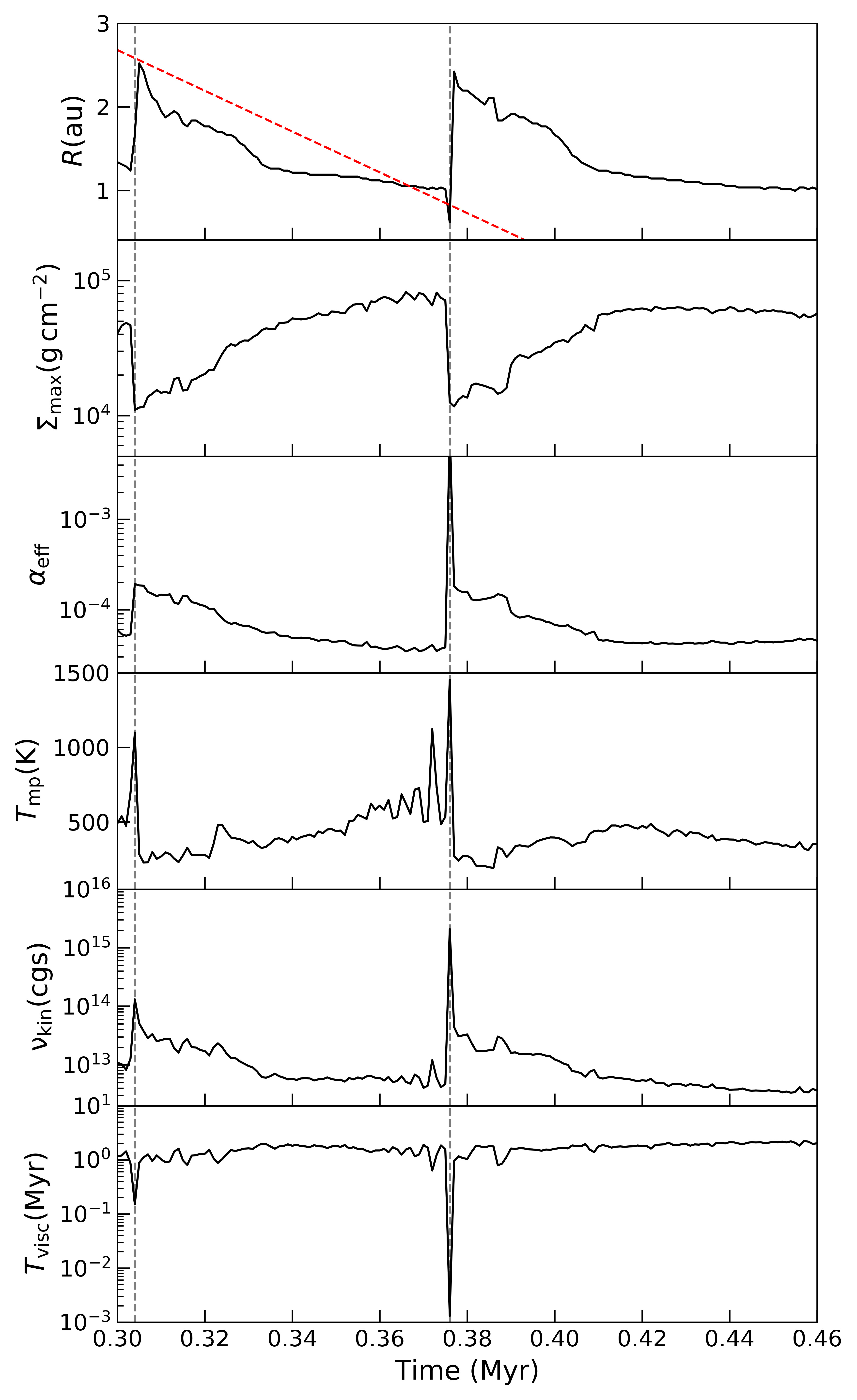}
}
\caption{
Properties of the most prominent ring in the fiducial layered disk model are plotted as it evolved.
The first panel shows the location of the most prominent ring, defined as the radius of the maximum surface density at a given time.
The remaining panels show $\Sigma_{\rm max}$, $\alpha_{\rm eff}$, $T_{\rm mp}$, kinematic viscosity and viscous timescale at this location. The vertical lines mark the time when the ring became MRI unstable and accreted onto the star. The diagonal red line in the first plot shows the average rate of migration.}
\label{fig:ringquant}
\end{figure}

With the understanding of the properties and dynamical nature of the inner disk, we can propose a coherent picture of how the rings form and evolve in the layered disk models.
As the disk evolves the gas does not accumulate at the outer edge of the dead zone because it undergoes a gradual transition, as well as the action of transient GI spirals across the dead zone.
When the gas is channeled the inner disk, the viscous torques work in the direction of reinforcing the pileup of material.
The positive feedback between viscous torques and accumulated gas produces the ring structures.
Since the rings are marginally gravitationally unstable, perturbations of the inner disk may give rise to GI fragmentation and spiral waves.
The associated transport of angular momentum results in the inward migration of the rings. 
The rings form within a few au from the inner edge of the dead zone, however, the exact location at a given time would depend of the interactions of all of the above factors.
The MRI instability is ultimately triggered in the disk and a ring rapidly depletes the accumulated material onto the star, resulting in an outburst.

}

\subsection{Implications for Planet Formation}
\label{subsec:planets}

\begin{figure*}
\hspace{3.5cm}  Azimuthal profiles    \hspace{2.7cm} \simname{model1\_T1300\_S100}  \hspace{0.8cm}  \simname{ model1\_const\_alpha}\\
\begin{tabular}{c}
\vspace{-0.22cm}\includegraphics[width=\textwidth]{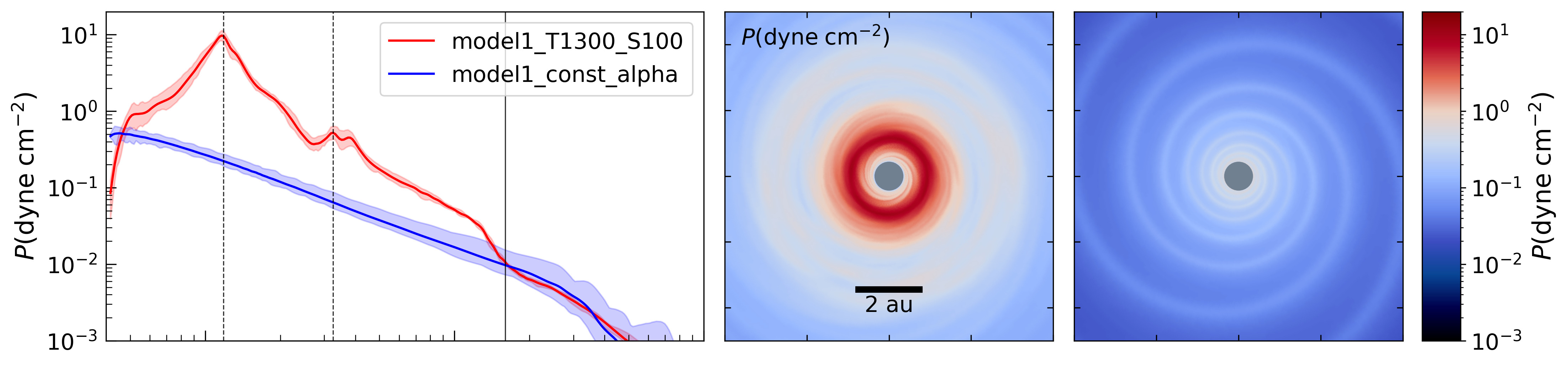} \\ 
\vspace{-0.24cm}\includegraphics[width=\textwidth]{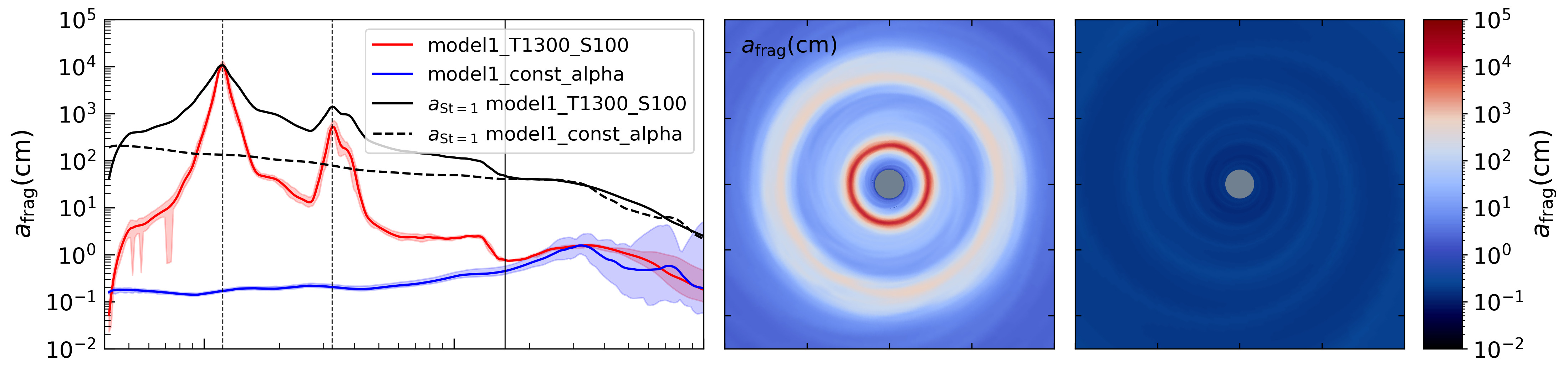} \\
  \includegraphics[width=\textwidth]{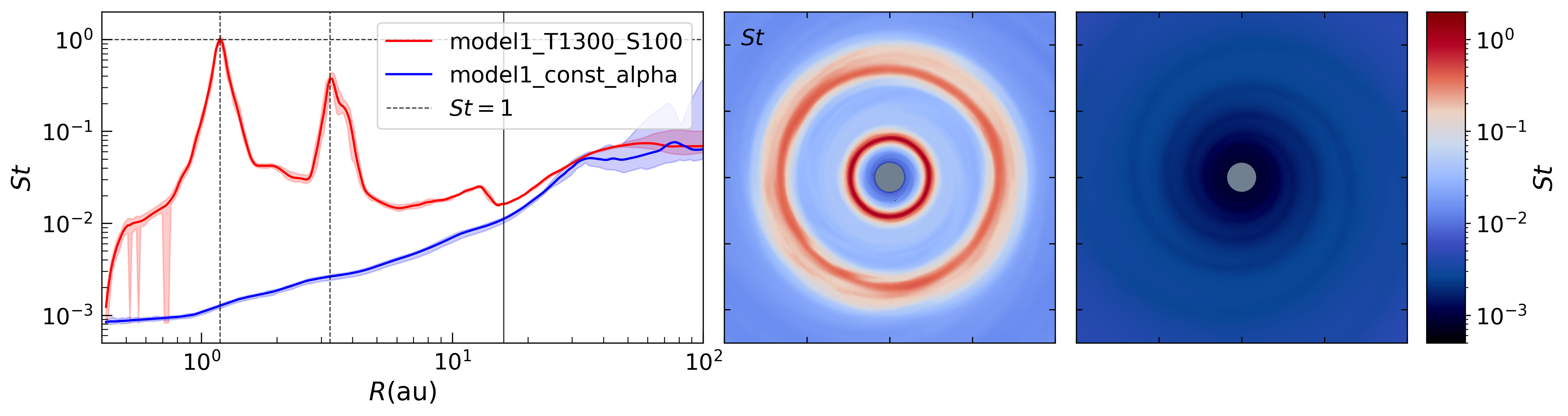} \\
\end{tabular}
\caption{Typical inner disk structure of both fiducial models is shown at 0.35 Myr, with respect to quantities that are relevant for planet formation --- vertically integrated pressure, {dust grain} fragmentation radius {(along with the grain size at Stokes number unity)} and Stokes number at the {dust grain} fragmentation radius. 
Left: Profiles of the azimuthally averaged parameters, with the shaded area showing the extent between the maximum and the minimum value at the given radius. The two vertical dashed lines mark the location of the two rings, while the solid line marks the outer extent of the dead zone. Right: The distribution of the same parameters is plotted in the inner $10 \times 10$ au box.}
\label{fig:planets}
\end{figure*}

The dynamical gaseous rings formed in a magnetically layered protoplanetary disk lie in the region which is of the greatest interest for terrestrial planet formation.
The dust grains which exist in the molecular cloud need to grow about 13 orders of magnitude in order to form planetesimals \citep{WC2011}.
In this section we investigate the implications of such ring structures on the accumulation and growth of dust.
In Figure \ref{fig:planets} we compare the inner disk structure of the fiducial layered disk model with that of the fully MRI active disk with respect to quantities that are relevant for the accumulation of dust --- vertically integrated pressure, {dust grain fragmentation radius} ($a_{\rm frag}$) and grain size at Stokes
number unity ($a_{St=1}$) as well as Stokes number at the {dust grain} fragmentation radius (St).
The leftmost column in this figure shows the azimuthal profiles, while on the right we compare the spatial distributions of these quantities in the inner $10\times 10$ au box. 
These quantities are calculated at 0.35 Myr, \ie, they correspond to the snapshot of the inner disk presented in Figure \ref{fig:inner1}.

Consider the first row of Figure \ref{fig:planets} which compares the vertically integrated pressure for the two fiducial models. 
The azimuthally averaged pressure for the fully MRI active model was monotonically decreasing with radial distance, although it showed a weak spiral structure within.
We know that dust particles experience a drag due to their relative velocity with respect to the gas in the disk in the direction of the pressure gradient (Weidenschilling 1977a).
Thus in the case of fully active disk, the dust is predicted to drift towards the star.
A weak spiral structure and associated local pressure maxima were shown to be inefficient in trapping dust particles in fully MRI-active disks, perhaps with an exception of the corotation region of the spiral pattern with the disk \citep{Vorobyov2018}.
In contrast, the layered disk model showed several maxima in the azimuthally averaged pressure near the location of the gaseous rings.
The dust grains can accumulate and grow in these regions of local pressure maxima.

The second row of Figure \ref{fig:planets} compares the {\rm dust grain fragmentation radius} for the two fiducial models, calculated as
\begin{equation}
    a_{\rm frag} = \frac{2 \Sigma v_{\rm frag}^2}{ 3 \pi \rho_{\rm s} \alpha_{\rm eff} c_{\rm s}^2},
\end{equation}
where the assumed values of fragmentation velocity is $v_{\rm frag}=10$~m~s$^{-1}$, and the internal density of the dust aggregate is $\rho_{\rm s} = 1.6$ g~cm$^{-3}$.
The {$a_{\rm frag}$} is a measure of the maximum size that the dust particles can achieve before they are destroyed through the process of fragmentation, which tends to limit the dust size in the inner disk \citep{Birnstiel2012,Vorobyov2018}.
{
The Stokes number for a particle of size $a$, under typical assumptions, is calculated as 
\begin{equation}
    {\rm St} = \frac{\Omega_{\rm k} \rho_{\rm s} a}{\rho c_{\rm s}},
\end{equation}
where the volume density, $\rho = \Sigma/ \sqrt{2 \pi} H$, $H$ being the local vertical scale height.
The Stokes number describes the aerodynamic properties of dust grains and thus the coupling of the dust to the gas flow.
The azimuthal plot of $a_{\rm frag}$ also shows the grain size at each radius that yields the Stokes number of unity.
Dust particles at this size will achieve the maximum radial drift velocity and therefore show the quickest radial migration.
Note that dust particles do not necessarily reach this size and barriers such as fragmentation, drift, bouncing etc. tend to limit their growth, unless these are overcome by means of other mechanisms, e.g. streaming \citep{Youdin2005} or gravitational \citep{Boss1997} instability.
}
In the case of fully MRI active disk, the dust particles encountered a fragmentation barrier when they were a few millimeter across. However, in the layered disk model the fragmentation barrier was much larger in comparison 
because of the large local surface density and low $\alpha_{\rm eff}$. 
Within the dead zone the maximum particle size remained mostly between few cm to 10 m, while occasionally approaching about 100 m in the vicinity of the rings. 

The third row in Figure \ref{fig:planets} shows the Stokes number of the dust particles at the {dust grain} fragmentation radius.
The small values of Stokes number for the fully MRI active disk mean that the dust particles were mostly coupled to and moved with the gas.
Since the dust concentration efficiency is highest for Stokes number of unity, the dust particles with this size ($a_{\rm frag}$) would accumulate very quickly (over the local orbital time scale) in the vicinity of the rings. 
Numerical studies of gas and dust evolution in a fully MRI active protoplanetary disk show that the radial inward drift of dust particles poses a significant barrier for planet formation \citep{Birnstiel2010,Vorobyov2018}.
The gaseous rings observed in our simulations could naturally form dust traps, effectively preventing this radial drift and providing sites for the accumulation of dust particles.
{However, the dynamic nature of the rings complicates the situation for their applicability for planet formation. 
As we saw in Section \ref{subsec:dynamical}, although the viscous timescale inside the rings was sufficiently long ($\approx 1$ Myr), the lifetime of the rings was much shorter -- only a few percent of this value -- before the MRI instability was triggered.
Evolution of the dusty component needs to be taken into account in order to estimate the timescale of the dust growth in this atypical environment and to determine whether the rings can successfully halt the radial drift long enough to form planetesimals. 
}

\subsection{Effects of Model Parameters}
\label{subsec:parameters}
So far we concentrated on the differences between a typical layered disk model and the classical approach of fully MRI active protoplanetary disk. 
As discussed in Section \ref{subsec:initial}, the parameters used within our layered disk model are constrained but not certain. 
The uncertainty in ${ T_{\rm crit}}$ arises due to the complexity of the dust and gas physics, and also that in estimating when the disk can be considered MRI active. 
On the other hand ${\Sigma_{\rm a}}$ may depend on the local conditions in the disk such as the flux of ionizing radiation, strength of magnetic field, as well as properties of dust grains.
In this section we investigate the dependence of the disk structure and evolution on these model parameters.
We present \simname{model1\_T1500\_S100} and \simname{model1\_T1300\_S10} to demonstrate the effects of changing critical temperature (${T_{\rm crit}=1500}$~K) and active layer thickness (${\Sigma_{\rm a}=10}$~g~cm$^{-2}$) respectively.
Most protostars have a mass less than 1 $\rm M_{\odot}$, so the dependence on mass is demonstrated with the help of \simname{model2\_T1300\_S100}.
In this case the initial cloud core mass was 0.346 $\rm M_{\odot}$, which is about 30\% of the mass of the core in the fiducial model.
As described in Table \ref{table:sims}, these models are otherwise identical to the fiducial layered disk model.
In the interest of conciseness the above three models are presented in detail, while the remaining simulations are referred to only when required.

\begin{figure}
\centering
\vspace{-0.15cm}\hspace{0.2cm} \includegraphics[width=7.7cm]{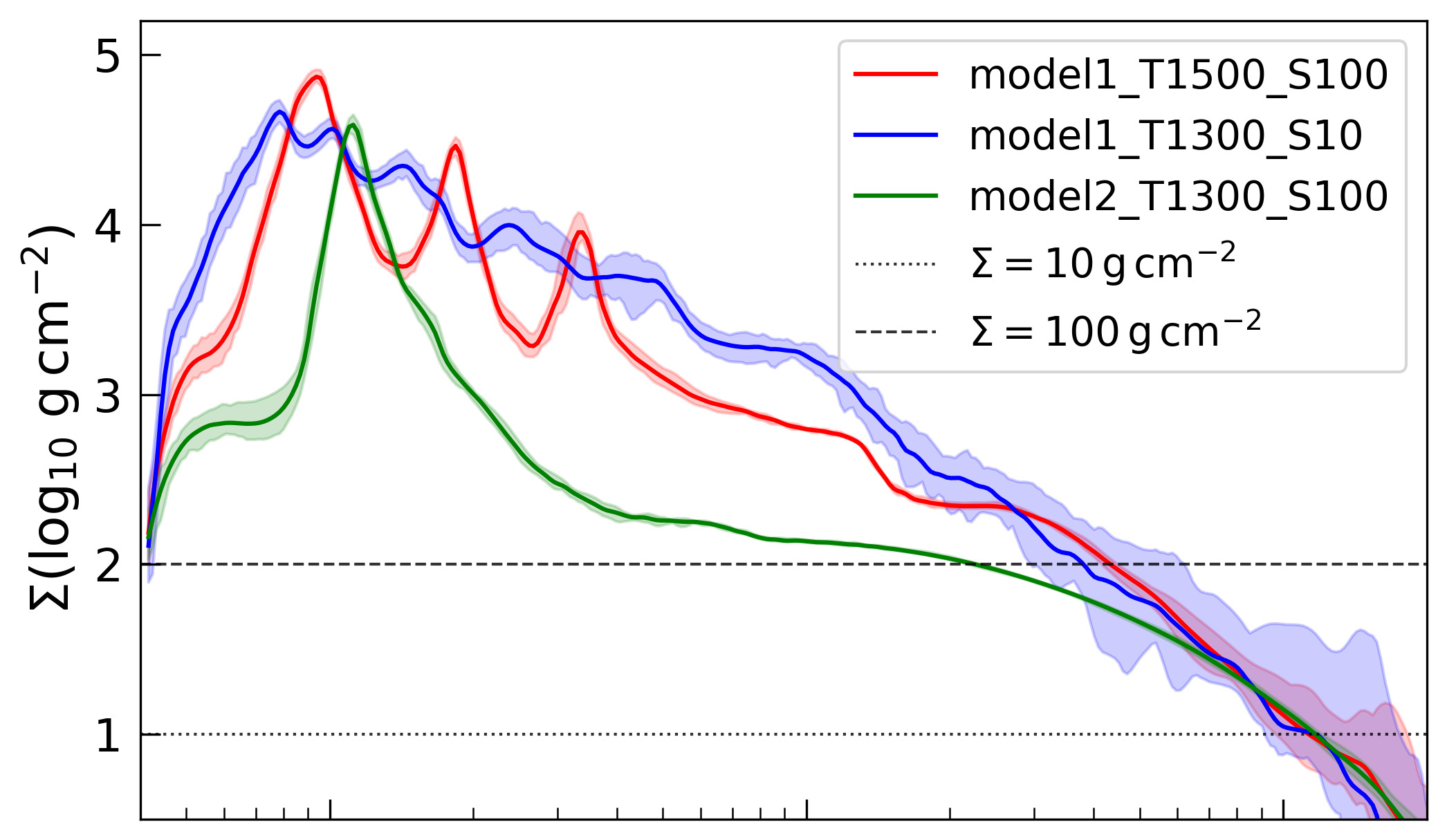} \\
\includegraphics[width=8cm]{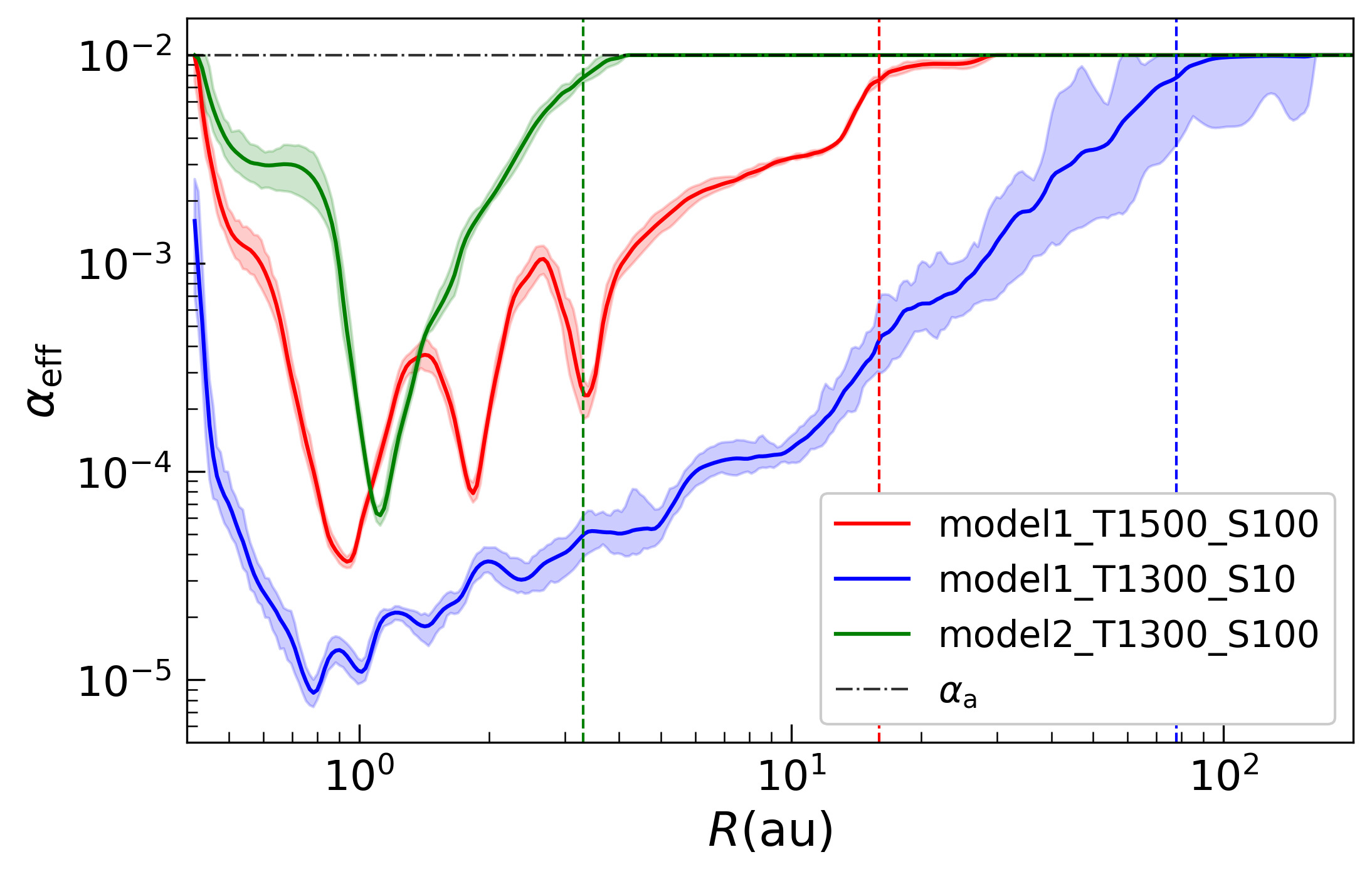} \\  
\caption{Typical inner disk structure of the fiducial adaptive alpha model is compared for the variation of $T_{\rm crit}$, $\Sigma_{\rm a}$, and mass of the parent core with respect to the distribution of gas surface density and $\alpha_{\rm eff}$ in the inner 200 au region. The solid lines show azimuthally averaged quantities, with the shaded area showing the extent between the maximum and the minimum value at the given radius. The color-coded vertical dashed lines in the second plot show the outer extent of the dead zone for the respective model.}
\label{fig:par1d}
\end{figure}

Figure \ref{fig:par1d} shows the typical disk structure in terms of azimuthal profiles of the three models --- \simname{model1\_T1500\_S100}, \simname{model1\_T1300\_S10} and \simname{model2\_T1500\_S100}.
We considered only $\Sigma$ and $\alpha_{\rm eff}$ as these two quantities are sufficient to demonstrate the salient properties of the rings as well as the dead zone.
Note that the abscissa shows the inner 200 au region in logarithmic units.
These stationary profiles are obtained at 0.35 Myr for the higher mass models (\simname{model1}), and at 0.18 Myr for the lower mass model (\simname{model2}).
Here we describe the main features of the typical ring structures, while the underlying causes between the differences are explained later in this section.
Consider \simname{model1\_T1500\_S100} (red lines) in Figure \ref{fig:par1d} showing the effects of increasing ${T_{\rm crit}}$, from 1300 K to 1500 K on the ring structures.
Three clearly formed rings can be seen in this case as compared to only two for the fiducial model at this time.
The maximum surface density as well as minimum in $\alpha_{\rm eff}$ at the location of the most prominent ring were similar to the fiducial model. 
The outer extent of the dead zone --- as shown by the red, dashed line in the second panel --- was about 16 au, almost equal to that of the fiducial model.  
The effect of changing ${\Sigma_{\rm a}}$ from 100 to 10 g~cm$^{-2}$ (\simname{model1\_T1300\_S10}, blue lines) was much more significant.
Although the maximum value of $\Sigma$ was not affected, a much deeper dead zone in terms of $\alpha_{\rm eff}$ was formed.
Unlike the other models with ${\Sigma_{\rm a}}=100$ g~cm$^{-2}$, the individual rings in this case were not as prominent and clearly distinguishable from the accumulated gas in the dead zone.
The variations between the maximum and the minimum at a given radius showed larger perturbations, indicating relatively more azimuthal asymmetry. 
Also note that several maxima in gas surface density and consequently in pressure were formed, which could trap dust at several radii throughout the dead zone.
The extent of the dead zone significantly increased to about 78 au, which is almost 5 times larger as compared to the fiducial model.
Consider \simname{model2\_T1500\_S100} (green lines) depicting the effect of lowering the mass of the parent core.
Only a single ring was formed at this time, while the maximum in $\Sigma$ or the minimum in $\alpha_{\rm eff}$ at the location of the ring were not significantly different as compared to the fiducial model.
The extent of the dead zone was much smaller, only 3.3 au, which is about 5 times smaller than the fiducial model. 
Note that for this lower mass model, the relatively flat surface density reached $\Sigma_{\rm a}=100$ g~cm$^{-2}$ at a much larger radius.

\begin{figure}
\centering
\simname{model1\_T1500\_S100}
\subfloat{
  \includegraphics[width=0.48\textwidth]{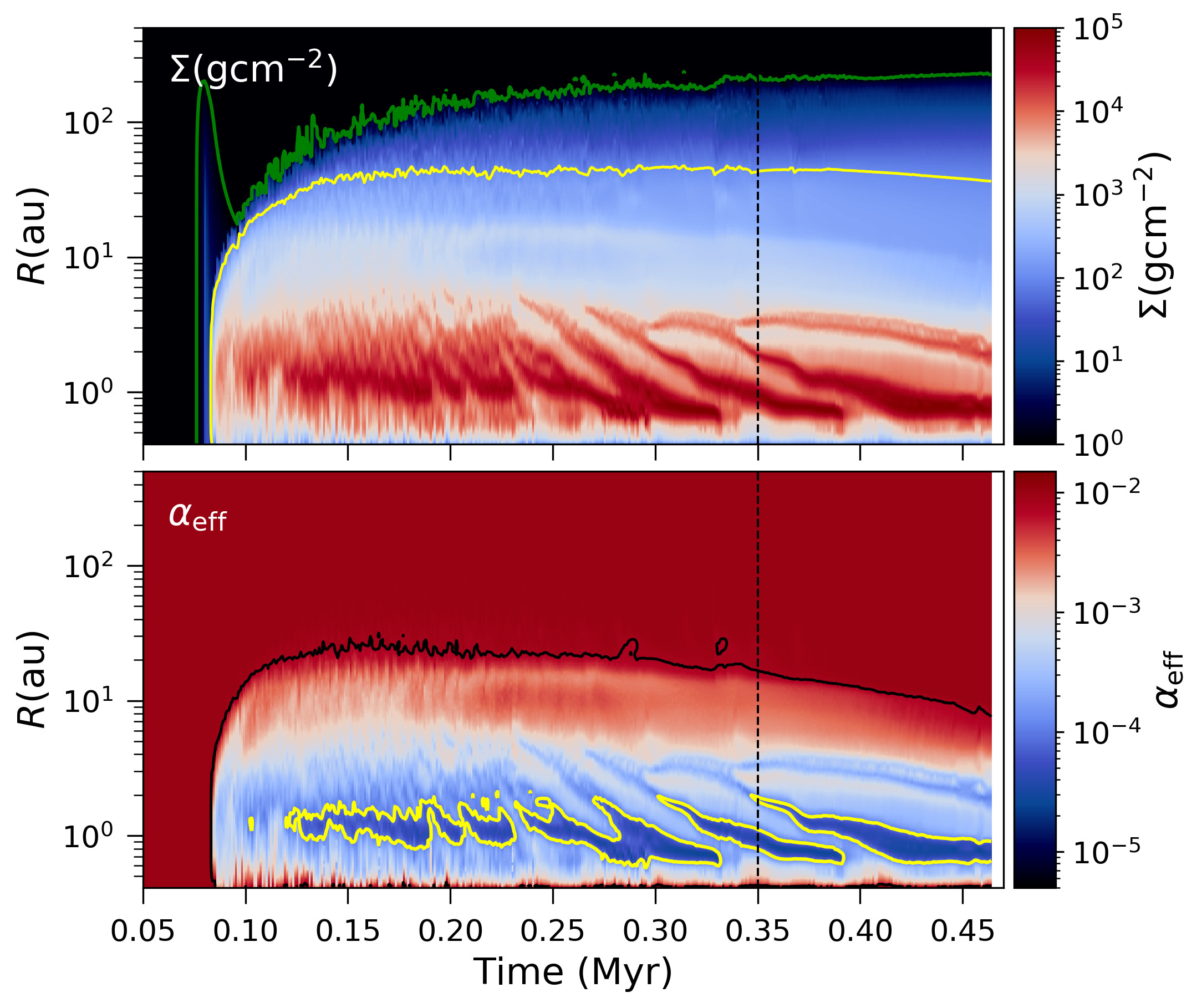}
}
\caption{The spacetime plots for \simname{model1\_T1500\_S100}, showing effects of increasing $T_{\rm crit}$ to 1500 K on the evolution of gas surface density and $\alpha_{\rm eff}$. Green contour in the first panel corresponds to $\Sigma=1$ g~cm$^{-2}$ and yellow contour to $\Sigma=\Sigma_{\rm a}=100 {\rm\, g\,cm^{-2}}$.
The black contour in the second panel shows the extent of the dead zone, while the yellow contour corresponds to $\alpha_{\rm eff}=10^{-4}$. The vertical dashed lines show the time corresponding to the azimuthal profiles in Figure \ref{fig:par1d}.}
\label{fig:parT1500}
\end{figure}

\begin{figure}
\centering
\simname{model1\_T1300\_S10}
\subfloat{
  \includegraphics[width=0.48\textwidth]{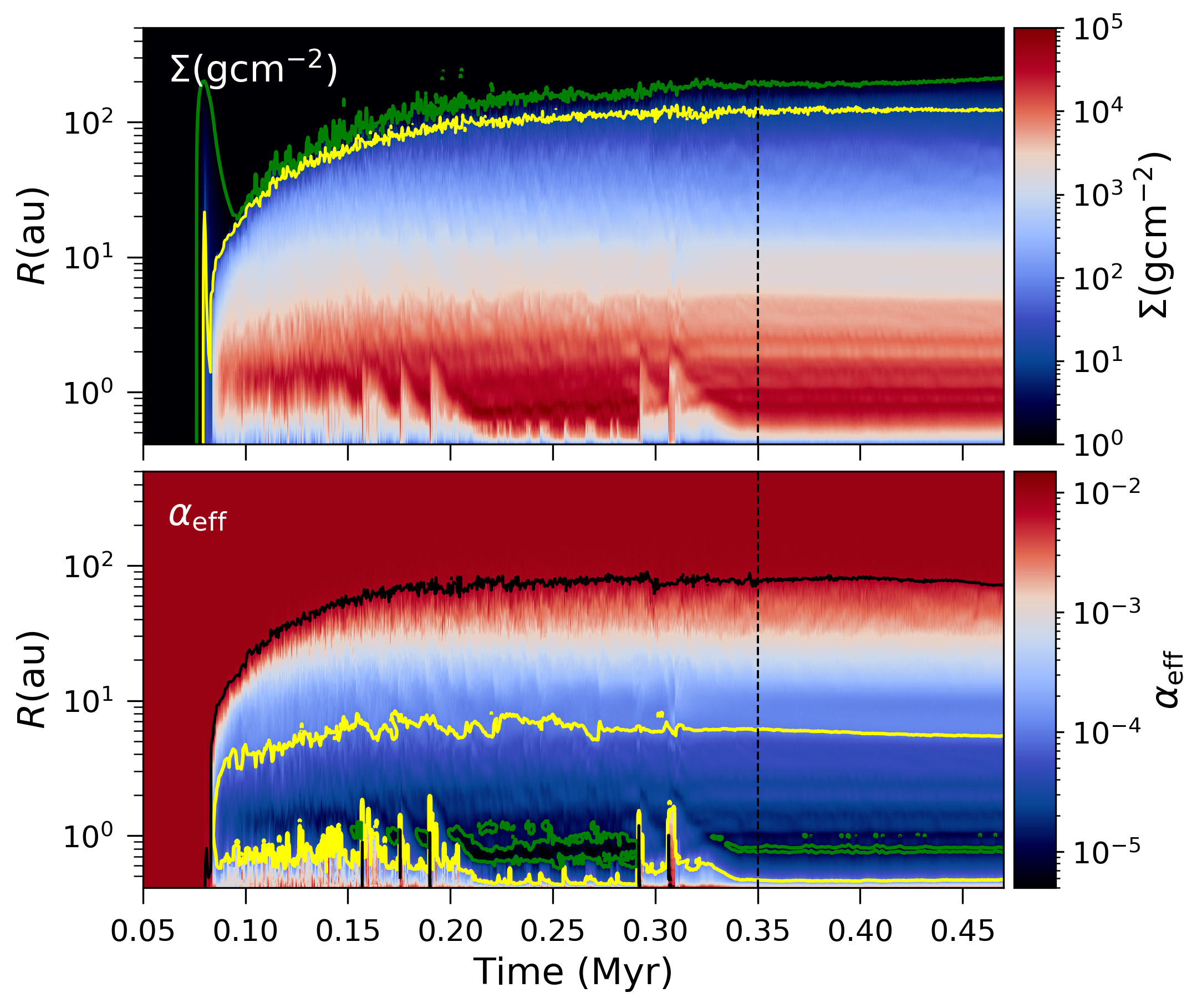}
}
\caption{The spacetime plots for \simname{model1\_T1300\_S10}, showing effects of decreasing $\Sigma_{\rm a}$ to $10 {\rm\, g\,cm^{-2}}$ on the evolution of gas surface density and $\alpha_{\rm eff}$. Green contour in the first panel corresponds to $\Sigma=1$ g~cm$^{-2}$ and yellow contour to $\Sigma=\Sigma_{\rm a}=10 {\rm\, g\,cm^{-2}}$.
The black contour in the second panel shows the extent of the dead zone, while the yellow and green contours correspond to $\alpha_{\rm eff}=10^{-4}$ and $10^{-5}$ respectively. The vertical dashed lines show the time corresponding to the azimuthal profiles in Figure \ref{fig:par1d}.}
\label{fig:parS10}
\end{figure}

\begin{figure}
\centering
\simname{model2\_T1300\_S100} 
\subfloat{
  \includegraphics[width=0.48\textwidth]{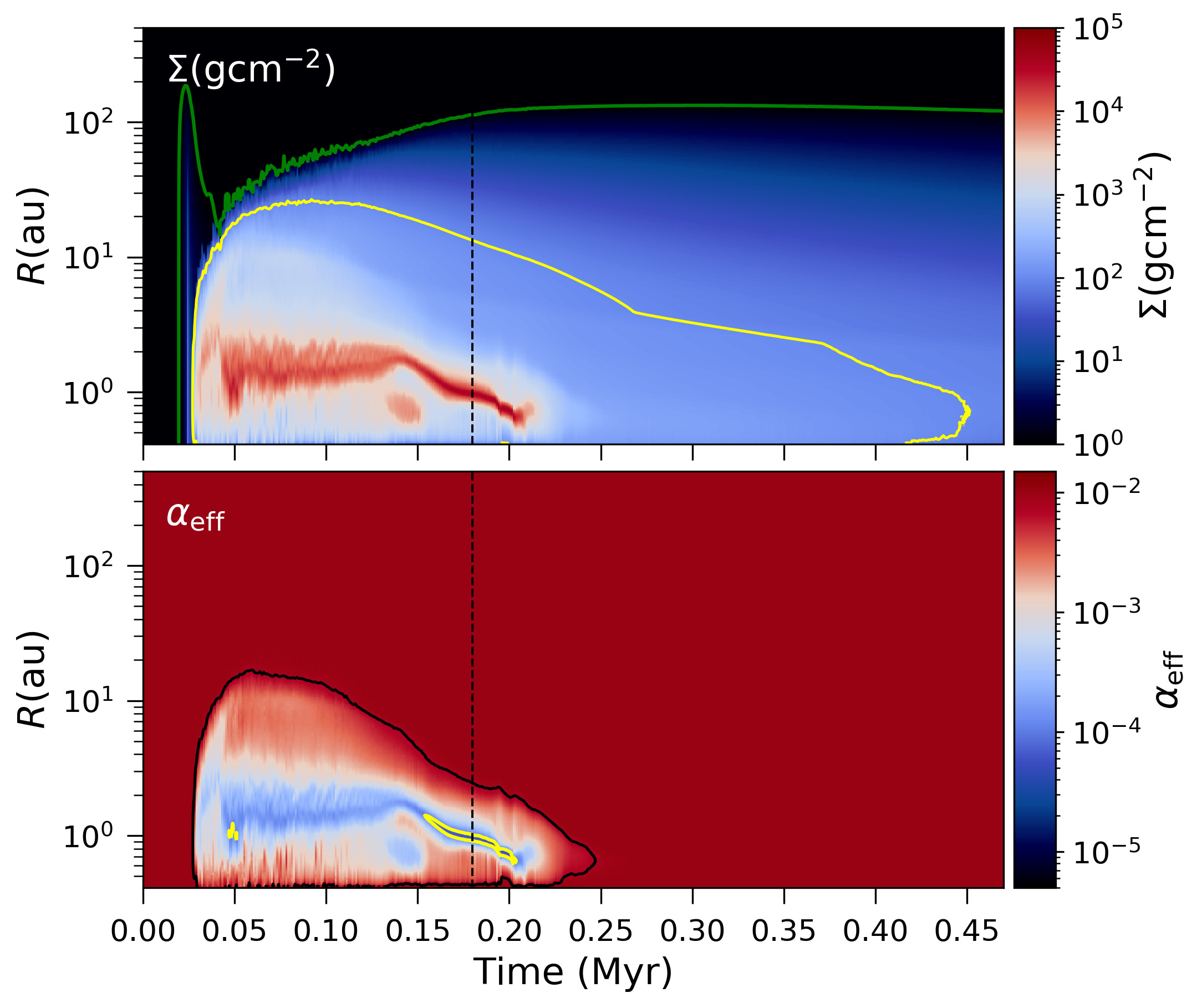}
}
\caption{The spacetime plots for \simname{model2\_T1300\_S100}, showing effects of a smaller cloud core mass on the evolution of gas surface density and $\alpha_{\rm eff}$. Green contour in the first panel corresponds to $\Sigma=1$ g~cm$^{-2}$ and yellow contour to $\Sigma=\Sigma_{\rm a}=100 {\rm\, g\,cm^{-2}}$.
The black contour in the second panel shows the extent of the dead zone, while the yellow contour corresponds to $\alpha_{\rm eff}=10^{-4}$. The vertical dashed lines show the time corresponding to the azimuthal profiles in Figure \ref{fig:par1d}.
}
\label{fig:parModel3}
\end{figure}

In order to demonstrate the effects of model parameters on the temporal evolution, the spacetime diagrams of the three simulations are shown in Figures \ref{fig:parT1500}-\ref{fig:parModel3}.
These figures are similar to Figure \ref{fig:spacetime1} for the fiducial models. 
Consider Figure \ref{fig:parT1500} for \simname{model1\_T1500\_S100}, showing the effects of increasing ${T_{\rm crit}}$ over about 0.5 Myr.
Notice that the qualitative behavior of the gaseous rings appears similar to the fiducial layered disk model.
The rings originated at the distance of a few au and migrated inwards which resulted in the diagonal patterns in the gas surface density.
The rate of inward migration was similar to the fiducial model.
The rings also disappeared suddenly, which indicates an instability similar to that described earlier resulting in FUor like accretion events.
The overall extent of the dead zone, denoted by the black contour in $\alpha_{\rm eff}$, was also similar to the fiducial model throughout the evolution. 
However, compared to the fiducial model, the ring structures display more complex behavior which was difficult to quantify. 
As shown earlier, more rings were formed simultaneously, \eg, between about 0.35--0.40 Myr, there are three rings in the disk as compared to two for the fiducial model.
Another example of complexity is that the two rings merged to form a single ring at about 0.25 Myr.
With the increase in ${ T_{\rm crit}}$, the MRI would be triggered at a higher temperature and the gas in the disk would remain in the disk for a longer time.
Since ${ T_{\rm crit}}$ was higher, the rings could reach a smaller radius before MRI instability was triggered.
Thus the inner rings were long-lived and gas moving in from the outer disk formed multiple outer rings.
The average time between MRI eruptions was about $55\,000$~yr, which is longer than the fiducial model, and thus consistent with late onset of the instability.
The complexity of the rings can also be seen in the spacetime diagram of $\alpha_{\rm eff}$. 
Thus we conclude that an increased ${ T_{\rm crit}}$ rendered the inner disk structure more complex in terms of the ring structures, however, the overall behavior was not affected significantly.  
We confirmed this conclusion with the comparison of the remaining three simulations with ${ T_{\rm crit}=1500}$K against their counterpart with ${ T_{\rm crit}=1300}$K (see Table \ref{table:sims}). 
This result can be expected since ${ T_{\rm crit}}$ is well constrained and it was increased by only about 15\%.

The spacetime diagrams for \simname{model1\_T1300\_S10} are presented in Figure \ref{fig:parS10} showing the effects of a decrease in the active layer thickness.
Consider the behavior of gas surface density in the first panel. 
Similar to the fiducial model the ring structures are formed in the inner disk at a few au radius.
As shown in Figure \ref{fig:par1d} these rings did not display a significant enhancement of gas surface density in the rings as compared to the surrounding material.
{ The major difference in the temporal behavior -- as compared to the previous simulations -- was that the 
rings showed $\sim 0.02$ Myr long diagonal patterns, and then no significant evolution.
Each of the inward migration was preceded by an MRI accretion event which was qualitatively similar to those described in Section \ref{subsec:dynamical}, and can be seen as the discontinuities in $\Sigma$ as well as jumps in $\alpha_{\rm eff}$ in Figure \ref{fig:parS10}.
Since the rings in \simname{model1\_T1300\_S10} were not as pronounced as in the previous cases, instead of complete disappearance of a ring, a partial ``cave in" of the inner disk region occurred.
During an MRI outburst the inner disk was rapidly depleted of material, and the newly formed ring migrated inward due to the action of GI spirals.
}
Notice that $\alpha_{\rm eff}$ fell below $10^{-5}$ in the vicinity of the rings, which was an order of magnitude lower as compared to the fiducial model.
{We conjecture that once the rings migrated inwards they were stable against fragmentation by inefficient cooling, and their slow evolution onward was a direct consequence of deepening of the rings with respect to $\alpha_{\rm eff}$.}
From Equation \ref{eq:alphaRD} it can be deduced that if $\Sigma_{\rm a}$ is decreased, $\alpha_{\rm rd}$ will also decrease accordingly. 
Thus lowering $\Sigma_{\rm a}$ by an order of magnitude resulted in a proportional decrease in $\alpha_{\rm eff}$ in the region of the rings.
The black contour in the second panel shows a much larger extent of the dead zone {(as compared to the simulations with $\Sigma_{a}=100 {\rm ~gcm^{-2}}$), about 78 au, which lasted throughout the 0.5 Myr of evolution.}
The width of the low viscosity region ($\alpha_{\rm eff}=10^{-4}$) was also increased with respect to the fiducial model because of {this accumulation of gas.}
As a consequence, the viscous timescale in the vicinity of the rings was of the order of 10 Myr, which is again 10 times larger than the fiducial model.
Due to the much wider dead zone as well as lower viscosity, the region must form a much more efficient bottleneck for mass and angular momentum transport as compared to the fiducial model.
{The higher surface gas density in the dead zone region possibly translated in strong GI spirals and a faster initial migration of the rings. 
These spirals may have resulted in a larger scatter between the maxima and minima seen in Figure \ref{fig:par1d}.}
The other three models with $\Sigma_{\rm a}=10$ g~cm$^{-2}$ from Table \ref{table:sims} evolved in a similar manner, showing a significantly slower evolution of the rings, as well as a much deeper and wider region with respect to $\alpha_{\rm eff}$.

Figure \ref{fig:parModel3} depicts the evolution of \simname{model2\_T1300\_S100}, where we show the effects of a decrease in the mass of the cloud core.
The change in the mass available for disk formation significantly affected the evolution of the inner disk region, which includes the gaseous rings as well as the dead zone.
Notice that with a smaller cloud core mass, the disk started forming at an earlier time at about 0.025 Myr, as compared to about 0.08 Myr for the higher mass models.
The spacetime diagram for $\Sigma$ shows that a ring started to form after some delay at about 0.15 Myr. 
As expected, the ring migrated inwards, however, this ``ring phase'' (approximately defined by the span of $\alpha_{\rm eff}=10^{-4}$ contour) did not last long, and the ring disappeared within only about 0.05 Myr.
As compared to the fiducial model this region of low $\alpha_{\rm eff}$, was much smaller in width as well.
We hypothesize that initially the ring formed because the gas surface density could be maintained above $\Sigma_{\rm a}$ due to the material moving in from the collapsing core through the outer disk.
Due to the cloud core being low mass, the outer disk contained a relatively limited amount of mass.
As the gas was transported inwards through the active layer, it was not replenished from the outer disk.
Thus a layered structure could not be maintained for an extended period of time.
{Despite of the low mass of the disk, we observed GI fragmentation and spirals in the rings. 
This may be because the low mass of the star increased dynamical time and the increased cooling efficiency was sufficient for the onset of GI.
}
The dead zone as depicted by the black contour also disappeared within a short time at about 0.25 Myr, and the disk essentially became fully MRI active thereafter.
Note that the yellow contour in the first panel showing $\Sigma=\Sigma_{\rm a}$ grossly overestimated the extent of the dead zone both in radius and time.
There were no MRI-triggered instabilities throughout the evolution, including when the ring was fully terminated. 
This indicates that the midplane temperature never reached $T_{\rm crit}$ in the relevant region and the ring eventually viscously dissipated. 
Thus we conclude that the mass of the initial cloud core can significantly affect the structure and evolution of the inner disk.

\begin{figure}
\centering
\simname{model2\_T1300\_S10}
\subfloat{
  \includegraphics[width=0.48\textwidth]{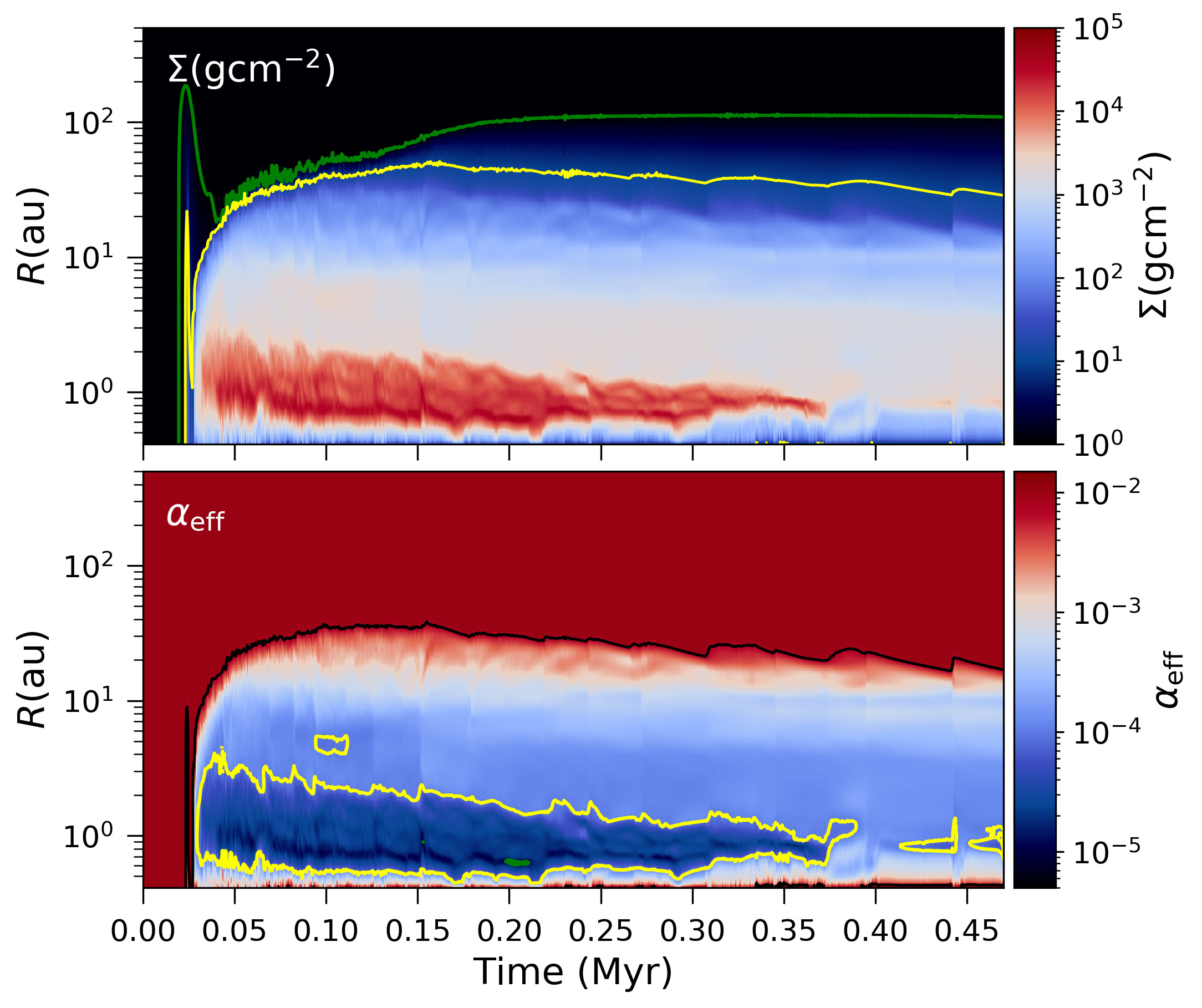}
}
\caption{The spacetime plots for \simname{model2\_T1300\_S10}, showing effects of a smaller cloud core mass as well as a decrease in $\Sigma_{\rm a}$ on the evolution of gas surface density and $\alpha_{\rm eff}$. Green contour in the first panel corresponds to $\Sigma=1$ g~cm$^{-2}$ and yellow contour to $\Sigma=\Sigma_{\rm a}=10 {\rm\, g\,cm^{-2}}$.
The black contour in the second panel shows the extent of the dead zone, while the yellow and green contours correspond to $\alpha_{\rm eff}=10^{-4}$ and $10^{-5}$ respectively.}
\label{fig:parModel3S10}
\end{figure}

In order to show the effect of ${\Sigma_{\rm a}}$ on the entire evolution of the ring phase, we investigate \simname{model2\_T1300\_S10}.
Figure \ref{fig:parModel3S10} depicts the spacetime diagrams for this model, showing the effect of a low active layer thickness on an initially low mass cloud core.
As expected, the dead zone formed was much wider and the rings formed were deeper in terms of $\alpha_{\rm eff}$, as compared to the analogous model with a higher $\Sigma_{\rm a}$ (\simname{model2\_T1300\_S100}).
However, the duration of the ring phase was several times longer in comparison.
This model also showed absence of MRI instabilities and associated outbursts, although the dead zone was maintained throughout the evolution. 
Thus we conclude that the inner disk structure as well as the properties of the rings were highly dependent on $\Sigma_{\rm a}$ as well as the initial cloud core mass, but only marginally depended on $T_{\rm crit}$.

\section{Conclusion}
\label{sec:summary}
In this paper we investigated the results of global protoplanetary disk formation simulations, starting with the collapse phase of the cloud core, which provided realistic initial conditions. 
The magnetically layered disk structure was modeled with an effective and adaptive $\alpha$ prescription. 
The inflow-outflow boundary conditions were implemented at the inner boundary of the computational domain, which was located at 0.4 au.
{As expected, we found that the layered disk models developed dead zones in the inner disk. 
However, due to the action of viscous torques, high surface density and low viscosity gaseous rings also formed within the innermost regions.} 
The rings were highly dynamical and migrated inwards towards the star, {primarily because of the angular momentum carried away through large--scale GI spirals.} 
The MRI was ultimately triggered resulting in a disk instability followed by an accretion event similar to FU Orionis eruptions. 

We contrasted the evolution of a representative layered disk model with that of a fully MRI active disk, starting with identical initial conditions.
Although on a large radii both of these models looked similar, the structure and behavior of the inner disk, at the scale of a few au, was significantly different. 
We found that dust particles with large fragmentation barrier (occasionally up to 100 m) could accumulate and rapidly grow inside the rings.  

We also studied the effects of the variation of the layered disk parameters, ${T_{\rm crit}}$ and ${\Sigma_{\rm a}}$, as well as the effect of a lower initial cloud core mass.
With an increase in ${T_{\rm crit}}$ from 1300 to 1500 K, the rings showed relatively mild effects such as more complex behavior and formation of multiple rings.
The effects of decreasing ${\Sigma_{\rm a}}$ from 100 to 10 g~cm$^{-2}$ were much more profound. 
The width of the dead zone as well as the depth of the rings in terms of $\alpha_{\rm eff}$ was increased by an order of magnitude, creating a more effective bottleneck for mass and angular momentum transport. 
As a result, the duration of the ring phase was longer by several times.
With a 60\% decrease in the cloud core mass (and a corresponding decrease in the final stellar mass), the ring phase was short-lived, and the disk also lacked MRI-triggered outbursts.

We saw that the gaseous rings develop pressure maxima where dust particles can get trapped rapidly, while the conditions within the rings are also favorable for a large fragmentation barrier.
The conventional core-accretion scenario of planet formation assumes an initial formation of nearly kilometer sized planetesimals \citep{GW1973, Pollack1996}. 
The ring structures can readily assist planet formation in the inner disk region.
The rings in our simulations had a limited lifespan and they ultimately became unstable through MRI triggering.
It is possible that the instabilities will not affect planetesimals, as large sized bodies should be decoupled from the gas and hence will not follow the accreting material during a disk instability.
The simulations showed that after the disk formation, stable rings appeared with some delay of about 0.1--0.2 Myr, depending on the mass of the parent core. 
Thus there may be a delay between disk formation and planet formation. 
The duration of the ring phase decreased significantly with a decrease in the mass of the parent core.
If the planet formation mechanism is related to these rings, the time available for this to occur can be significantly reduced with the stellar mass. 
The lower mass stars should also undergo much less episodic accretion through FUor--like outbursts, which can affect the chemistry of the disk. 
Protoplanetary disk having strong magnetic environment or powerful winds may have an advantage with respect to planet formation, as the inner disk is better shielded against cosmic rays. 
The number of planet cores formed may increase with this shielding, as both duration and depth of the rings (with respect to $\alpha_{\rm eff}$) was increased for a smaller magnetically active layer.

Our simulations predict that with a magnetically layered disk structure, gaseous rings are formed ubiquitously during the T Tauri phase of a low mass star.
Although these rings displayed a remarkable contrast as compared to a fully MRI active disk, they are contained within the innermost regions of the disk, up to only a few au. 
Observing such small distances offers significant challenges even to state of the art facilities.
\Eg, typical resolving capability of Atacama  Large  Millimeter/submillimeter  Array (ALMA) at the nearest star forming region of Taurus molecular cloud is about 25 milliarcsecond or 3.5 au \citep{alma2015}, which is not sufficient for resolving the inner disk structure.
However, with upcoming radio interferometry facilities such as Next Generation Very Large Array \citep[ngVLA,][]{Carilli2015} or by combining observations from several instruments \eg, Multi AperTure mid-Infrared SpectroScopic Experiment (MATISSE) and ALMA \citep{Kobus2019}, it is possible to probe and constrain the disk structure in this region and observationally verify our results.  

At this point we would like to mention one major limitation of our model --- we did not consider the self--consistent evolution of the dust component, which dominates the opacity of a protoplanetary disk, and hence the emission properties.
Processes such as growth of dust grains, their drift relative to the gas, dust self--gravity and back reaction can affect the environment in the vicinity of the rings. 
However, building a comprehensive model is challenging due to the complexity of the physics involved and running such simulations is generally computationally expensive \citep[\eg, see][]{Haworth2016}.
In the future we plan to conduct simulations of magnetically layered protoplanetary disks using a hydrodynamic model similar to \cite{Vorobyov2018}, which includes evolution of a two-part dusty component.

\begin{acknowledgements}
      {We thank the anonymous referee for constructive comments which improved the quality of the paper significantly.}
      This project has received funding from the European Research Council (ERC) under the European Union's Horizon 2020 research and innovation programme under grant agreement No 716155 (SACCRED).
      Eduard Vorobyov acknowledges support from the Russian Science Foundation grant 17-12-01168.
      Part of this work was supported by the German
      \emph{Deut\-sche For\-schungs\-ge\-mein\-schaft, DFG\/} project number Ts~17/2--1.
      We also acknowledge support from the Hungarian OTKA Grant No. K-119993.
\end{acknowledgements}



\end{document}